\documentclass[3p]{elsarticle}

\usepackage{amssymb}
\usepackage{amsmath}

\usepackage{color,xcolor}

\journal{}
\begin{document}

\begin{frontmatter}
\title{Normal form computation of nonlinear dispersion relationship for locally resonant metamaterial}
\author[1,2,3]{Tao Wang}
\author[3]{Cyril Touzé\corref{cor1}}
\ead{cyril.touze@ensta.fr}
\author[1,2]{Haiqin Li}
\author[1,2]{Qian Ding\corref{cor1}}
\ead{qding@tju.edu.cn}

\cortext[cor1]{Corresponding author}

\affiliation[1]{organization = {Department of Mechanics, Tianjin University},
city = {Tianjin},
postcode = {300072},
country = {China}}
\affiliation[2]{organization = {Tianjin Key Laboratory of Nonlinear Dynamics and Control},
city = {Tianjin},
postcode = {300072},
country = {China}}
\affiliation[3]{organization = {IMSIA, ENSTA-CNRS-EDF, Institut Polytechnique de Paris},
city = {Paris},
postcode = {91762 Palaiseau Cedex},
country = {France}}
\begin{abstract}
   This article is devoted to the application of the parametrisation method for invariant manifold with a complex normal form style (CNF), for the derivation of high-order approximations of underdamped nonlinear dispersion relationships for periodic structures, more specifically by considering the case of a locally resonant metamaterial chain incorporating damping and various nonlinear stiffnesses.
   Two different strategies are proposed to solve the problem. In the first one, Bloch’s assumption is first applied to the equations of motion, and then the nonlinear change of coordinates provided by the complex normal form style in the parametrisation method is applied. This direct procedure, which applies first the wave dependency to the original physical coordinates of the problem, is referred to as CNF-BP (for CNF applied with Bloch's assumption on physical coordinates). In the second strategy, the nonlinear change of coordinates provided by the parametrisation method, which relates the physical coordinates to the so-called normal coordinates, is first applied. Then the periodic assumption is used, thus imposing a Bloch wave ansatz on the normal coordinates. This method will be referred to as CNF-PN (for CNF with a periodic assumption on normal coordinates).
   In the conservative case, the CNF-PN strategy exhibits superior capability in capturing complex wave propagation phenomena, whereas the CNF-BP strategy encounters limitations in handling non-fundamental harmonics and the nonlinear interactions between host oscillators. The influence of truncation order on the accuracy of CNF-PN is further examined, demonstrating its effectiveness in extending the validity limit. For underdamped systems, the CNF-PN is rigorously validated and systematically compared against numerical techniques, a classical analytical perturbation technique (the method of multiple scales), and direct numerical time integration of annular chain structures. The results confirm the exceptional accuracy of the CNF-PN in predicting nonlinear dispersion relationships, damping ratios, invariant manifolds, and wave attenuation characteristics, as long as the validity limit of the asymptotic expansion is not reached. This advancement provides a novel and efficient analytical and numerical tool for studying nonlinear metamaterials.
\end{abstract}
\end{frontmatter}

\section{Introduction}
\label{sec:1}
Nonlinear metamaterials~\cite{RN1,FronkReviewer} refer to artificially designed periodic structures with controllable nonlinear elements, such as geometric nonlinearity~\cite{RN2,RN3,RN4}, impact~\cite{RN5,RN6}, quasi-zero stiffness~\cite{RN7,RN8} and bistable configurations~\cite{RN9,RN10,RN11}. When including the micro-resonator into the lattice, a subclass of metamaterial can be achieved, known as locally resonant metamaterial~\cite{RN1,FronkReviewer}. Due to the coupling effects between lattices, host-resonator interactions, and amplitude-dependent dispersion characteristics, both general and locally resonant metamaterial exhibit unique wave manipulation characteristics, such as adaptivity~\cite{RN12,RN13}, bandgap broadening~\cite{RN8,RN14,RN15,RN16}, non-reciprocal transmission~\cite{RN17,RN18,RN19}, higher harmonic generation~\cite{RN3,RN4,RN15,RN20,RN21,RN22}, and internal resonances~\cite{RN11,RN23}, making them highly promising for applications in acoustics, mechanical vibration control, and elastic waveguides. However, compared to their linear counterpart, the complex nonlinear dynamics and the inherent infinite degrees of freedom (DOFs) present significant theoretical challenges in accurately predicting dispersion properties, nonlinear damping effects, and other nonlinear behaviours. Hence, there is a need to develop an efficient and reliable theoretical framework for analysing such systems.

The primary challenge in analysing nonlinear metamaterials arises from the infinite number of DOFs in an oscillator chain, which renders the direct application of conventional nonlinear analysis techniques difficult. To address this issue, certain assumptions~\cite{FronkReviewer} are typically introduced to constrain the infinite-dimensional model within a finite-sized domain, thereby enabling further analysis. The most common approach exploits the periodicity of the lattice structure to directly apply Bloch’s assumption on the physical model~\cite{RN24}. This assumption relates the variables of adjacent cells by stating that the dynamics are given by a propagative wave. If the vector $\mathbf{x}_n$ describes the displacement of cell $n$, Bloch's assumption leads to writing the displacement vectors of the two adjacent cells as: $\mathbf{x}_{n\pm 1} = e^{\mp jk}\mathbf{x}_n$, where
$j$ and $k$ respectively stand for the imaginary unit and the wavenumber. Thanks to this assumption, the infinite problem is replaced by a finite-dimensional dynamical model as a function of the wavenumber $k$. This strategy facilitates the extraction of nonlinear dispersion characteristics using conventional nonlinear analysis techniques~\cite{RN2,RN6,RN16,RN25,RN26,RN27,RN28}. For instance, Fang {\it et al.}~\cite{RN25} adopted this approach to simplify the modelling of a nonlinear diatomic chain and compared the accuracy of perturbation methods, the harmonic balance method (HBM), and the homotopy method in predicting nonlinear dispersion relationships. Similarly, Gong {\it et al.}~\cite{RN27} combined the HBM with Bloch’s periodic assumption to investigate the evolution of bandgaps in strongly nonlinear triatomic acoustic metamaterials. Moreover, several studies have integrated Bloch’s assumption with Hamiltonian perturbation theory~\cite{RN29} and the method of multiple scales (MMS)~\cite{RN26} to compute invariant manifolds in locally resonant metamaterials. For continuous systems, Zhao {\it et al.}~\cite{RN2} incorporated Bloch’s periodic assumption into a transfer function framework to examine the nonlinear dispersion properties of metamaterial beams with geometric nonlinear oscillators. Likewise, Shen {\it et al.}~\cite{RN16,RN28} employed Bloch’s assumption in conjunction with the Galerkin method to reduce the dynamical complexity of locally resonant metamaterial beams and plates, enabling the prediction of their dispersion solutions via the MMS.

The strategy of directly applying Bloch’s periodic assumptions to physical coordinates for simplifying the dynamic model offers advantages such as computational efficiency and compatibility with existing nonlinear analysis techniques for finite structures. However, this approach has inherent limitations. Since Bloch’s assumption 
is fundamentally based on the Bloch wave ansatz~\cite{RN24}, 
it only accounts for a part of the full wave expression. While very effective for linear problems or uncoupled lattices thanks to the ensuing simplifications of exponential Bloch wave terms, it overlooks critical components when nonlinearities induce some couplings between adjacent lattices, hence severely limiting its nonlinear applicability. Moreover, Bloch’s periodic assumption fails to accurately capture the emergence of non-fundamental harmonics, leading to significant errors in cases such as metamaterials with quadratic nonlinearities. To address these limitations, Settimi {\it et al.} embedded Bloch’s assumptions within a multiple scales analysis framework~\cite{RN30}, applying them at each time scale to mitigate errors associated with directly imposing this assumption on the physical model. This approach effectively predicted the dispersion properties and invariant manifolds of metamaterials incorporating quadratic and cubic nonlinearities in internal resonators. A more advanced strategy involves directly incorporating a complete Bloch wave ansatz into the solution process~\cite{RN11,RN31,RN32,RN33,RN34,Manktelow2011,Narisetti2D}. Within an asymptotic analysis framework, Narisetti {\it et al.}~\cite{Narisetti2D,RN35} applied perturbation methods to investigate one- and two-dimensional periodic nonlinear lattices. They derived dispersion relations and explored the influence of nonlinearity on wave propagation by embedding a Bloch wave formulation at each perturbation order. Manktelow {\it et al.}~\cite{Manktelow2011} further integrated the MMS with Bloch wave solutions and compared its performance against the Lindstedt–Poincaré perturbation method in capturing wave-wave interactions. Yi \textit{et al.}~\cite{RN4} also combined the MMS with a Bloch wave ansatz to analyse nonlinear chains featuring axial–rotational coupling, demonstrating the capability of geometric nonlinearity to generate higher harmonics. Beyond asymptotic approaches, the HBM can also leverage the Bloch wave ansatz to circumvent the limitations of Bloch’s assumption in nonlinear settings. Narisetti {\it et al.}~\cite{RN33} assumed that the response of a periodic lattice could be expressed as a superposition of waves with integer multiples of the fundamental frequency and wavenumber, enabling the analysis of dispersion properties in one- and two-dimensional strongly nonlinear periodic lattices governed by Hertzian contact laws via HBM. More recently, the HBM has been used in conjunction with damped nonlinear normal modes approximated with the extended periodic motion concept (EPMC)~\cite{Krack_EPMC}, to develop a numerical framework tailored for weakly damped, strongly nonlinear metamaterials~\cite{RN11}. This approach was applied to a metamaterial chain with bistable resonators, revealing intricate internal resonance behaviours and wave attenuation characteristics.

At present, the analysis of nonlinear metamaterials primarily relies on numerical HBM~\cite{RN11,RN27,RN33} and analytical asymptotic theories (\emph{e.g.}, perturbation techniques~\cite{RN25,RN35}, such as  MMS~\cite{RN8,RN22,RN23,RN30,RN31,RN32}). However, these methods have inherent limitations when applied to infinite periodic structures: numerical techniques are difficult to provide analytical insights into the underlying nonlinear dynamic behaviours and dispersion characteristics, while traditional perturbation approaches are typically restricted to low-order approximations due to computational complexity, limiting their accuracy. Additionally, existing studies predominantly focus on conservative systems or weakly damped scenarios~\cite{RN11,RN31,RN32}, lacking an accurate and robust technique to investigate the interplay between damping and nonlinear coupling effects.

Among perturbative techniques, normal form theory plays a special role since its first introduction by Poincaré and Dulac~\cite{RN39,RN40}. In the field of vibration, it has been used by many different investigators to derive approximate solutions~\cite{Nayfeh:NF,RN38} or make the link with the concept of Nonlinear Normal Modes (NNMs)~\cite{Jezequel91,touze03-NNM,TOUZE:JSV:2006}. In recent years, the introduction of the parametrisation method for invariant manifold~\cite{RN41} has significantly enhanced the practical applicability of normal form theory. In particular, the parametrisation method embeds in the same framework two classical methods used in model order reduction, namely the center manifold theorem and the normal form approach~\cite{RN41,RN36}; and shows that both techniques can be interpreted as different styles of parametrisation. The formalism also allows for efficient and automatable high order approximations, that have been used successfully for model order reduction~\cite{RN42,PONSIOEN2018,RN43,RN45,Vizzaccaro2024}. In vibration theory, the complex normal form (CNF) style~\cite{RN43,RN46,RN37} has the advantage of providing analytical expressions for the backbone curve~\cite{BreunungHaller18}. In particular, the CNF can be advantageously used to compute symbolic and numerical solutions with arbitrary orders, allowing systematic study of {\em e.g.} softening/hardening behaviour, super/subharmonic resonances, and parametric vibrations in single- and two-degree-of-freedom systems, as reported for instance in~\cite{RN46}. However, its extension to infinite periodic structures remains unexplored, particularly in addressing nonlinear lattice interactions, host-resonator nonlinearities, and damping-nonlinearity coupling effects.

In this study, an infinite locally resonant metamaterial chain with damping and various nonlinear stiffnesses is considered in Section~\ref{sec:2}, as an example to demonstrate how the normal form procedure can cope with such problems and provide a high-order approximation for the nonlinear dispersion relationships.
Two different strategies are introduced in Section~\ref{sec:3}: (i) the CNF-BP applies Bloch’s assumption directly onto the physical coordinates, and (ii) the CNF-PN enforces periodic constraints (from Bloch wave expression) onto the normal coordinates, which describe the dynamics on the related invariant manifold. By integrating these strategies within the direct parametrisation method for invariant manifold (DPIM)~\cite{RN43,Vizzaccaro2024}, and using the CNF style, the nonlinear dispersion curves are obtained with arbitrary order truncations. The performance of both strategies is systematically compared in undamped scenarios (Section~\ref{sec:4.1}), revealing the advantages of the CNF-PN strategy in capturing host nonlinearity, zero-order harmonic, higher-order harmonics, and complex wave propagation phenomena. The effect of truncation order on prediction accuracy is also examined in Section~\ref{sec:4.2}. For damped systems (Section~\ref{sec:4.3}), numerical integration results from a finite annular chain are used to validate the accuracy of CNF-PN in predicting nonlinear dispersion relations, damping ratios, and wave attenuation characteristics. Additionally, comprehensive comparisons with EPMC-HBM and MMS are conducted, highlighting the strengths of the proposed approach. This study establishes a novel paradigm for the theoretical analysis of infinite periodic nonlinear systems and advances the applicability range of normal form theory.

\section{Infinite metamaterial chain with damping and nonlinear stiffnesses}
\label{sec:2}
\subsection{Dynamic model}
\label{sec:2.1}
\begin{figure}
    \centering
    \includegraphics[width=0.9\linewidth]{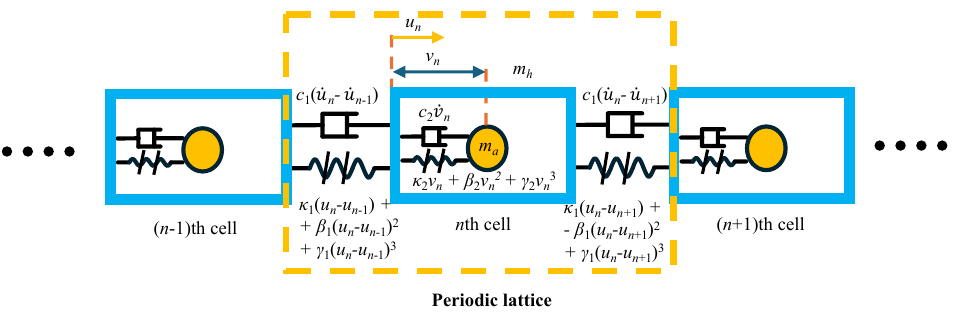}
    \caption{Schematic diagram of the infinite locally resonant metamaterial chain with damping and various nonlinear stiffnesses.}
    \label{f:1}
\end{figure}

The present study aims at adapting the direct parametrisation method to derive the normal form for infinite metamaterial structures. The resulting CNF can be used to derive analytical expressions for the nonlinear dispersion relationship and damping ratio. To illustrate the proposed methodology, an infinite nonlinear metamaterial chain, shown in Fig.~\ref{f:1}, is considered as a representative example, where damping, quadratic and cubic nonlinear stiffnesses are included in the internal attachments and adjacent lattices. The governing equation of $n$-th lattice is given by
\begin{equation}
    \label{eq:1}
    \mathbf{M}\ddot{\mathbf{x}}_n + \sum_{i=-1}^{1}{\mathbf{C}_i\dot{\mathbf{x}}_{n+i}}+ \sum_{i=-1}^{1}{\mathbf{K}_i\mathbf{x}_{n+i}} +\mathbf{F}_n\left(\mathbf{x}_{n}\right)+\mathbf{F}_{nb,-1}\left(\mathbf{x}_{n}-\mathbf{x}_{n-1}\right) +\mathbf{F}_{nb,+1}\left(\mathbf{x}_{n}-\mathbf{x}_{n+1}\right) = 0,
\end{equation}
where
\begin{subequations}
\begin{align}
\mathbf{M} &= \begin{bmatrix}
m_h+m_a & m_a\\
m_a & m_a
\end{bmatrix},\mathbf{K}_{\pm1} = \begin{bmatrix}
    -\kappa_1 & 0\\
    0 & 0
\end{bmatrix},\mathbf{K}_0 = \begin{bmatrix}
    2\kappa_1 & 0\\
    0 & \kappa_2
\end{bmatrix}, 
\mathbf{C}_{\pm1} = \begin{bmatrix}
    -c_1 & 0\\
    0 & 0
\end{bmatrix},
\mathbf{C}_0 = \begin{bmatrix}
    2c_1 & 0\\
    0 & c_2
\end{bmatrix},\\
\mathbf{x}_{n+i} &= \begin{bmatrix}
    u_{n+i}\\
    v_{n+i}
\end{bmatrix}, 
\mathbf{F}_{nb,\pm1} = \begin{bmatrix}
    \mp\beta_1 (u_n-u_{n\pm1})^2 + \gamma_1 (u_n-u_{n\pm1})^3\\
    0
\end{bmatrix}, \mathbf{F}_{n} = \begin{bmatrix}
    0\\
    \beta_2 v_n^2 + \gamma_2 v_n^3
\end{bmatrix}.
\end{align}
\end{subequations}
Here, $m_h, \kappa_1$, and $c_1$ represent the mass, linear stiffness, and damping of the host oscillators, respectively, while $m_a, \kappa_2$, and $c_2$ correspond to the mass, linear stiffness, and damping of the internal resonators. The variables $u_n$ and $v_n$ denote the displacement of the $n$-th host oscillator and the relative displacement between the host mass block and its internal attachment. The coefficients $\beta_1, \beta_2$ and $\gamma_1, \gamma_2$ characterise the quadratic and cubic nonlinear stiffnesses that arise from interactions between adjacent lattices and between the host mass block and its internal resonator. The notations $\dot{()}$ and $\ddot{()}$ denote the first, $\mathrm{d}/\mathrm{d}t$, and second derivatives, $\mathrm{d}^2/\mathrm{d}t^2$, with respect to time, $t$.

\subsection{Linear dispersion relationships}
\label{sec:2.2}
The linear dispersion solutions are the key characteristics to explain the behaviour of propagative waves in such a metamaterial lattice. Besides, they are used in perturbative solutions as the first-order term from which the perturbations brought about by the nonlinearities will be searched for. Consequently, their derivation is the common first step of analysis. The DPIM, as formulated for example in~\cite{RN43,RN45,Vizzaccaro2024}, uses a first-order formulation such that Eq.~\eqref{eq:1} is rewritten as:
\begin{subequations}
    \label{eq:2}
    \begin{align}
    \begin{aligned}
    \mathbf{M}\dot{\mathbf{y}}_n + \sum_{i=-1}^{1}{\mathbf{C}_i\dot{\mathbf{x}}_{n+i}} +\sum_{i=-1}^{1}{\mathbf{K}_i\mathbf{x}_{n+i}}+\mathbf{F}_{n,2}(\mathbf{x}_n,\mathbf{x}_n)+\mathbf{F}_{n,3}(\mathbf{x}_n,\mathbf{x}_n,\mathbf{x}_n)
    +\mathbf{F}_{nb,2}(\hat{\mathbf{x}}_{-1},\hat{\mathbf{x}}_{-1})\\+\mathbf{F}_{nb,3}(\hat{\mathbf{x}}_{-1},\hat{\mathbf{x}}_{-1},\hat{\mathbf{x}}_{-1})-\mathbf{F}_{nb,2}(\hat{\mathbf{x}}_{+1},\hat{\mathbf{x}}_{+1})+\mathbf{F}_{nb,3}(\hat{\mathbf{x}}_{+1},\hat{\mathbf{x}}_{+1},\hat{\mathbf{x}}_{+1})=0,
    \end{aligned}\\
    \mathbf{M}\dot{\mathbf{x}}_n = \mathbf{M}\mathbf{y}_n,
    \end{align}
\end{subequations}
where
\begin{subequations}
\begin{align}
\mathbf{y}_{n} &= \dot{\mathbf{x}}_n,\hat{\mathbf{x}}_{\pm1} = \mathbf{x}_n-\mathbf{x}_{n\pm1},\mathbf{F}_{n,2} = \begin{bmatrix}
    0\\
    \beta_2 \mathbf{x}_{n2}^2
\end{bmatrix}, \\\mathbf{F}_{n,3} &= \begin{bmatrix}
    0\\
    \gamma_2 \mathbf{x}_{n2}^3
\end{bmatrix}, \mathbf{F}_{nb,2} = \begin{bmatrix}
    \beta_1\hat{\mathbf{x}}_{\pm11}^2\\
    0
\end{bmatrix},\mathbf{F}_{nb,3} = \begin{bmatrix}
    \gamma_1\hat{\mathbf{x}}_{\pm11}^3\\
    0
\end{bmatrix},
\end{align}
\end{subequations}
where $\mathbf{x}_{ni},\hat{\mathbf{x}}_{\pm1i}$ implies the $i$-th element in $\mathbf{x}_{n}$ and $\hat{\mathbf{x}}_{\pm1}$. The displacements $\hat{\mathbf{x}}_{\pm1}$ and $\mathbf{x}_{n\pm1}$ are variables shared by adjacent lattices.
To derive a solution from Eqs.~\eqref{eq:2}, an assumption must be imposed to constrain these shared variables. With the purpose of deriving the linear dispersion curves, all the nonlinear terms are discarded from Eq.~\eqref{eq:2}, and the Bloch's periodic condition, yielding   $\mathbf{x}_{n\pm1} = e^{\mp jk}\mathbf{x}_{n},\hat{\mathbf{x}}_{\pm1} = (1-e^{\mp jk})\mathbf{x}_{n}$, where $k$ denotes the wavenumber, is introduced.
Substituting these relations into Eqs.~\eqref{eq:2} and neglecting the nonlinear terms, the system reduces to a linear eigenvalue problem:
\begin{equation}
    \label{eq:5}
    \Lambda(k)\mathbf{B}(k)\mathbf{Y}(k) = \mathbf{A}(k)\mathbf{Y}(k),
\end{equation}
where
\begin{subequations}
\begin{align}
\mathbf{K}(k) &= \mathbf{K}_0 + 2\cos{k}\mathbf{K}_{-1},\mathbf{C}(k)=\mathbf{C}_0+2\cos{k}\mathbf{C}_{-1},\\
\mathbf{B} &= \begin{bmatrix}
\mathbf{C}(k)&\mathbf{M}\\
    \mathbf{M}&\mathbf{0}
\end{bmatrix},\mathbf{A} =\begin{bmatrix}
    -\mathbf{K}(k)&\mathbf{0}\\
    \mathbf{0}&\mathbf{M}
\end{bmatrix}.
\end{align}
\end{subequations}
When $k\neq0$, Eq.~\eqref{eq:5} possesses two eigensolutions with complex conjugate eigenvalues. Let us denote, for $i=1,\cdots,4$, $\Lambda_i(k)$ the eigenvalues, and $\mathbf{Y}_i(k)$ the corresponding eigenvectors. Importantly, all the solutions depend on the wavenumber $k$. When no ambiguity is at hand, the dependence on $k$ is omitted to make the notations lighter. In addition, owing to the symmetry of the matrices $\mathbf{B}(k)$ and $\mathbf{A}(k)$, the left and right eigenvectors coincide and are both represented by $\mathbf{Y}_i(k)$. It is customary to distinguish the acoustic branch from the optic branch. In the present case, the two complex conjugate eigenvalues $(\Lambda_1,\Lambda_2)$, where $\Lambda_2=\bar{\Lambda}_1$, refer to the acoustic branch with eigenvectors $\mathbf{Y}_1$, and $\mathbf{Y}_2=\bar{\mathbf{Y}}_1$. On the other hand, the pair $(\Lambda_3,\Lambda_4)$, with $\Lambda_4=\bar{\Lambda}_3$, denote the optic branch, with eigenvectors $\mathbf{Y}_3$, and $\mathbf{Y}_4=\bar{\mathbf{Y}}_3$.

According to the relationship, $\mathbf{y}_n = \dot{\mathbf{x}}_n$, the eigenvectors take the form $\mathbf{Y}_i = [\boldsymbol{\phi}_i;\Lambda_i\boldsymbol{\phi}_i]$, with eigenvalues given by $\Lambda_1=-\zeta_a\omega_a+j\omega_a\sqrt{1-\zeta_a^2}$ and $\Lambda_3=-\zeta_o\omega_o+j\omega_o\sqrt{1-\zeta_o^2}$. Here, $(\zeta_a,\zeta_o)$ refers to the damping ratio of the acoustic and optic branches, $(\omega_a,\omega_o)$ refers to the frequency, while $(\boldsymbol{\phi}_1,\boldsymbol{\phi}_3)$ indicates the complex waveforms associated with the second-order eigenvalue problem in Eq.~\eqref{eq:1}, and are represented in the form of column vectors. As customary done in eigenvalue problems, a normalisation constraint on the eigenvectors is also introduced, $\mathbf{Y}_{i}^\mathrm{T}\mathbf{B}\mathbf{Y}_{s} = \delta_{is}$, where $\delta_{is}$ stands for the Kronecker symbol ($\delta_{is} = 1$ when $i=s$, and $\delta_{is} = 0$ when $i\neq s$). It should be noted that the normalisation condition is only valid for $k\neq0$ or $k=0$ in the case of the optic branch.

When $k = 0$, Eq.~\eqref{eq:5} produces two identical eigensolutions for the acoustic branch due to the singularity of the stiffness matrix $\mathbf{K}(0)$, specifically, $\Lambda_1=\Lambda_2 = 0$ and $\mathbf{Y}_1 =\mathbf{Y}_2=[1;0;0;0]$, which indicates that the host oscillators undergo rigid-body motion without velocity or relative displacement between the host mass block and the internal attachment. For the optic case, whether $k=0$ or $k\neq0$, the $\Lambda_3$ and $\mathbf{Y}_3$ yield distinct but conjugate values with respect to $\Lambda_4$ and $\mathbf{Y}_4$.

The particular solution with a rigid-body mode when $k=0$, will have some important consequences when deriving the nonlinear solutions. Indeed, since the normal form theory makes large use of the commensurability relationships between the eigenvalues, special care will be devoted to the treatment of the vanishing eigenvalues.

\section{Direct parametrisation method for invariant manifold in metamaterial}
\label{sec:3}

This section is devoted to the application of the normal form approach embedded in the DPIM to obtain nonlinear dispersion curves with arbitrary order of accuracy. Two strategies are provided to handle the shared variables, $\hat{\mathbf{x}}_{\pm1}$ and $\mathbf{x}_{n\pm1}$, in Eqs.~\eqref{eq:2}, thereby adapting the DPIM to infinite nonlinear metamaterial chains. The first and most conventional approach directly applies Bloch’s periodic condition to the physical coordinates to constrain the shared variables. This strategy has been widely employed in nonlinear metamaterial studies~\cite{RN25,RN26,RN27,RN28}. However, this approach introduces several limitations, including its inability to effectively address the generation of waves with zero-order or higher-order frequencies and wavenumbers, as well as its inadequacy in handling the nonlinearity present in host oscillators. The second strategy, which represents the significant contribution of this work, involves imposing periodic constraints (based on Bloch wave ansatz~\cite{RN11,RN31,RN32}) on the normal coordinates of the CNF. This approach effectively mitigates the drawbacks associated with the direct application of Bloch’s assumption, providing a more robust and efficient framework for the implementation of CNF in infinite nonlinear periodic structures. 

As a side note, one can remark that the two different strategies differ in the order of using Bloch's assumption and the nonlinear change of coordinates provided by the DPIM and the normal form. The choice of first deriving the nonlinear change of coordinate, and then applying the periodicity assumption on the temporal solution of the normal coordinates, can be compared with analysis led on nonlinear oscillation with the normal form reported in~\cite{NeildNF00,RN38,Wagg2019,RN46}. Indeed, in these cases, a single harmonic solution is introduced in the normal coordinates, and all the nonlinearities appearing in the solution in physical coordinates, are generated through the nonlinear mapping. The CNF-PN technique proposed in Section~\ref{sec:3.2} shares similarities with these treatments.

It is also worth mentioning that both strategies considered in this work are limited to cases without internal resonance between different dispersion solutions. Accordingly, only the acoustic or optical linear eigenvectors are selected as the basis for computing the nonlinear parts of the CNF and invariant manifold. The direct application of Bloch’s assumption cannot correctly capture non-fundamental harmonics and is therefore inadequate for handling internal resonances induced by wave–wave interactions. In contrast, the CNF-PN approach adopts a complete Bloch wave ansatz to accurately capture the resonance relationships among different wave modes. Such relationships have also been reported in references~\cite{RN22,RN23,Manktelow2011} regarding wave-wave internal resonance. While the CNF-PN method shows promise in addressing wave–wave interactions, the presence of internal resonances introduces significant complexity. This article is thus devoted to handling the case without internal resonance, whereas the case with internal resonance is left for future works.


\subsection{CNF-BP Strategy: direct application of Bloch's assumption to physical coordinates}
\label{sec:3.1}
For the CNF-BP strategy, Bloch's assumption is first employed to constrain the physical coordinate and set $\mathbf{x}_{n\pm1} = e^{\mp jk}\mathbf{x}_{n},\hat{\mathbf{x}}_{n\pm1} = (1-e^{\mp jk})\mathbf{x}_{n}$, allowing Eqs.~\eqref{eq:2} to be rewritten as
\begin{subequations}
    \label{eq:3}
    \begin{align}
    \mathbf{M}\dot{\mathbf{y}}_n + \mathbf{C}(k)\dot{\mathbf{x}}_{n} + \mathbf{K}(k)\mathbf{x}_{n}+\mathbf{G}(\mathbf{x}_n,\mathbf{x}_n;k)+\mathbf{H}(\mathbf{x}_n,\mathbf{x}_n,\mathbf{x}_n;k)
    =0,\\
    \mathbf{M}\dot{\mathbf{x}}_n = \mathbf{M}\mathbf{y}_n,
    \end{align}
\end{subequations}
where
\begin{subequations}
\begin{align}
\mathbf{G}(\mathbf{x}_n,\mathbf{x}_n;k) &= \mathbf{F}_{n,2}(\mathbf{x}_n,\mathbf{x}_n)+((1-e^{jk})^2-(1-e^{-jk})^2)\mathbf{F}_{nb,2}(\mathbf{x}_{n},\mathbf{x}_{n}),\\
\mathbf{H}(\mathbf{x}_n,\mathbf{x}_n,\mathbf{x}_n;k) &= \mathbf{F}_{n,3}(\mathbf{x}_n,\mathbf{x}_n,\mathbf{x}_n) + ((1-e^{jk})^3+(1-e^{-jk})^3)\mathbf{F}_{nb,3}(\mathbf{x}_{n},\mathbf{x}_{n},\mathbf{x}_{n}).
\end{align}
\end{subequations}
By treating $k$ as a parameter, Eqs.~\eqref{eq:3} can be regarded as a nonlinear dynamic equation with a finite size, as highlighted in the previous Section~\ref{sec:1}. Consequently, the DPIM as formulated for example in~\cite{RN43,RN46,Vizzaccaro2024}, can be directly applied to Eqs.~\eqref{eq:3}. To derive the nonlinear dispersion curve related to either the acoustic or the optic mode, two nonlinear mappings, relating the displacement $\mathbf{x}_n$ and its velocity $\mathbf{y}_n$ of $n$-th lattice, to a two-dimensional normal coordinate $\mathbf{z}$, are introduced as:
\begin{subequations}
\label{eq:4}
\begin{align}
\mathbf{x}_n &= \mathbf{\Psi}(\mathbf{z})=\mathbf{\Phi}\mathbf{z} + \sum_{p=2}^{o}\sum_{q=1}^{q_p}\mathbf{\Psi}^{(p,q)}\mathbf{z}^{\boldsymbol{\alpha}(p,q)},\\ 
\mathbf{y}_n &= \mathbf{\Upsilon}(\mathbf{z})=\mathbf{\Phi}\boldsymbol{\lambda}\mathbf{z}_ + \sum_{p=2}^{o}\sum_{q=1}^{q_p}\mathbf{\Upsilon}^{(p,q)}\mathbf{z}^{\boldsymbol{\alpha}(p,q)}.
\end{align}
\end{subequations}
In these equations, $\mathbf{\Phi}$ and $\boldsymbol{\lambda}$ consist of the eigenvectors and eigenvalues (along with their conjugates) obtained from the linear dispersion analysis in Section~\ref{sec:2.2}. Specifically, they are constructed using $\mathbf{\Phi} = [\boldsymbol{\phi}_{1},\boldsymbol{\phi}_{2}]$ and $\boldsymbol{\lambda} = \mathrm{diag}([\Lambda_{1},\Lambda_{2}])$ for acoustic case or replacing the subscript "1, 2" with "3, 4" for optic branch. The normal coordinate $\mathbf{z} = [z_1;z_2]$ is two-dimensional and Eq.~\eqref{eq:4} describes the geometry of the related invariant manifold. Using the multi-index notation~\cite{Reed1980,RN41,Vizzaccaro2024}, $\mathbf{z}^{\boldsymbol{\alpha}(p,q)}$ represents an order-$p$ monomial, expressed as $z_1^{\alpha_1}z_2^{\alpha_2}$, where $\boldsymbol{\alpha}(p,q) = [\alpha_1;\alpha_2]$, $\sum_{s=1}^{2}{\alpha_s(p,q)} = p$ the polynomial degree, and $\alpha_1,\alpha_2$ are non-negative integers. In the summations, $o$ refers to the maximal order of the asymptotic expansion, while $q_p$ denotes the total number of monomials of order $p$, equal here to $p+1$ since $\mathbf{z}$ is two-dimensional. Finally, the auxiliary index $q$ in the summations is introduced to differentiate between the monomials with the same order $p$, following a given order, see {\em e.g.}~\cite{Reed1980,Vizzaccaro2024}.

In conjunction with Eqs.~\eqref{eq:4}, the reduced dynamics on the invariant manifold is introduced as:
\begin{equation}\label{eq:reduceddyna}
    \dot{\mathbf{z}} = \boldsymbol{\lambda}\mathbf{z} + \sum_{p=2}^{o}\sum_{q=1}^{q_p}\mathbf{f}^{(p,q)}\mathbf{z}^{\boldsymbol{\alpha}(p,q)}.
\end{equation}
All the introduced coefficients in the nonlinear mappings and reduced dynamics, namely $\mathbf{\Psi}^{(p,q)},\mathbf{\Upsilon}^{(p,q)}$, and $\mathbf{f}^{(p,q)}$, represent the coefficient vectors associated with the monomial $\mathbf{z}^{\boldsymbol{\alpha}(p,q)}$ in the displacement and velocity mappings of invariant manifolds, as well as in the reduced dynamics. Substituting these expressions into Eqs.~\eqref{eq:3} and equating the coefficients of each monomial yields the so-called homological equations of order $p$~\cite{RN41}, that needs to be solved recursively. The first-order problem is automatically satisfied due to the selection of $\mathbf{\Phi}$ and $\boldsymbol{\lambda}$ based on linear dispersion solutions. Regarding order-$p$ problems, they are solved iteratively. 

The general derivation of the order-$p$ homological equation can be found in~\cite{RN41}, and its application to vibrating systems for example in~\cite{RN43,RN45,Vizzaccaro2024}. Interestingly, these equations can be solved for each monomial separately, easing the generalisation of automated arbitrary order solutions. For the sake of brevity, the details of this calculation are reported in Section~\ref{sec:3.2} and~\ref{sec:Add}. Only the equation related to the monomial $\mathbf{z}^{\boldsymbol{\alpha}(p,q)}$ is given here. It reads:
\begin{equation}
    \label{eq:6}
    \begin{array}{l}
    \left(\sigma^{(p,q)}\mathbf{B}(k)+\mathbf{A}(k)\right)\left[\begin{array}{l}
    \boldsymbol{\Psi}^{(p,q)}\\
    \boldsymbol{\Upsilon}^{(p,q)}
    \end{array}\right] +\sum_{s=1}^{2} f_{s}^{(p,q)}\mathbf{B}(k) \left[\begin{array}{l}
    \boldsymbol{\Phi}_{s} \\
    \boldsymbol{\Phi}_{s}\lambda_{s}
    \end{array}\right]\\
    =\left[\begin{array}{c}
    -\mathbf{C}(k)\boldsymbol{\mu}^{(p,q)}-\mathbf{M} \boldsymbol{\nu}^{(p,q)}-\mathbf{G}^{(p,q)}-\mathbf{H}^{(p,q)} \\
    -\mathbf{M} \boldsymbol{\mu}^{(p,q)}
    \end{array}\right],
    \end{array}
\end{equation}
where $\lambda_s$, $f_{s}^{(p,q)}$, and $\boldsymbol{\Phi}_{s}$ denote elements in the $s$-th row or column of $\mathrm{diag}(\boldsymbol{\lambda})$, $\boldsymbol{f}^{(p,q)}$, and $\boldsymbol{\Phi}$, respectively. 
The remaining variables are defined as follows

\begin{equation}
\label{eq:7}
    \left\{\begin{array}{l}
    \sigma^{(p,q)}=\sum_{s=1}^{2}\alpha_s(p,q)\lambda_s,\\
    \boldsymbol{\mu}^{(p,q)}=\sum_{s=1}^{2} \sum_{p_1=2}^{p-1} \sum_{q_1=1}^{q_{p1}} {\alpha_s(p-p_1+1,q_2)\boldsymbol{\Psi}^{(p-p_1+1,q_2)} f_{s}^{(p_1,q_1)}},\\
    \boldsymbol{\nu}^{(p,q)}=\sum_{s=1}^{2} \sum_{p_1=2}^{p-1} \sum_{q_1=1}^{q_{p1}}{\alpha_s(p-p_1+1,q_2)\boldsymbol{\Upsilon}^{(p-p_1+1,q_2)} f_{s}^{(p_1,q_1)}},\\
    \mathbf{G}^{(p,q)}=\sum_{p_1=1}^{p-1} \sum_{q_1=1}^{q_{p1}}{ \mathbf{G}\left(\boldsymbol{\Psi}^{(p_1,q_1)}, \boldsymbol{\Psi}^{(p-p_1,q_2)};k\right)},\\
    \mathbf{H}^{(p,q)}=\sum_{p_1=1}^{p-2} \sum_{p_2=1}^{p-k-1} \sum_{q_1 = 1}^{q_{p1}} \sum_{q_2 = 1}^{q_{p2}}\mathbf{H}\left(\boldsymbol{\Psi}^{(p_1,q_1)}, \boldsymbol{\Psi}^{(p_2,q_2)}, \boldsymbol{\Psi}^{(p-p_1-p_2,q_3)};k\right).
    \end{array}\right.
\end{equation}
where
\begin{equation}
\left\{\begin{array}{l}
    \boldsymbol{\alpha}(p-p_1+1,q_2) = \boldsymbol{\alpha}(p,q)-\boldsymbol{\alpha}(p_1,q_1) + \boldsymbol{e}_s, \\
    \boldsymbol{\alpha}(p-p_1,q_2) = \boldsymbol{\alpha}(p,q) - \boldsymbol{\alpha}(p_1,q_1),
    \boldsymbol{\alpha}(p-p_1-p_2,q_3) = \boldsymbol{\alpha}(p,q) - \boldsymbol{\alpha}(p_1,q_1) - \boldsymbol{\alpha}(p_2,q_2).
    \end{array}\right.
\end{equation}
The unit vector $\boldsymbol{e}_s$ is introduced as $\boldsymbol{e}_s= [1;0]$, when $s=1$, or $\boldsymbol{e}_s= [0;1]$ when $s = 2$. Notably, in the summations appearing in Eqs.~\eqref{eq:7}, combinations of the vectors describing the monomial powers appear as: $\boldsymbol{\alpha}(p-p_1+1,q_2), \boldsymbol{\alpha}(p-p_1,q_2), \boldsymbol{\alpha}(p-p_1-p_2,q_3)$. Importantly, these may yield negative elements and contradict the definition of the multi-index notation. Such cases are excluded during the calculation process of Eqs.~\eqref{eq:7}. In Eq.~\eqref{eq:6}, the variables on the left-hand side—$\boldsymbol{\Upsilon}^{(p,q)}$, $\boldsymbol{\Psi}^{(p,q)}$, and $f_{s}^{(p,q)}$—are unknown at the $p$-th order, while the right-hand side terms arise from lower-order solutions. Since the number of unknowns exceeds the number of equations, the system is underdetermined, necessitating the introduction of a parametrisation style to solve the underdeterminacy~\cite{RN41}. In the present derivation, the complex normal form (CNF) style is used as it is related to the contention of the simplest reduced-order model, {\it i.e.} the normal form~\cite{RN41,RN37}. Interestingly, after a polar coordinate transformation, the CNF can produce analytic formulations for computing the nonlinear frequency and instantaneous damping ratio at a given $k$~\cite{BreunungHaller18,RN46}.
To select the CNF style, the vector $\tilde{\mathbf{Y}}_m$ is introduced and defined as $\tilde{\mathbf{Y}}_m = \mathbf{Y}_m$ for acoustic branches or $\tilde{\mathbf{Y}}_m = \mathbf{Y}_{2+m}$ for optical branches, where $m = 1, 2$. The $\mathbf{Y}_m$ denotes the eigenvectors obtained from the linear dispersion analysis in Section~\ref{sec:2.2}. Depending on the targeted dispersion branch, the appropriate form of $\tilde{\mathbf{Y}}_m=\mathbf{Y}_m$ or $\mathbf{Y}_{m+2}$ is selected. Then, leveraging the normalisation condition $\mathbf{Y}_i^\mathrm{T} \mathbf{B} \mathbf{Y}_s = \delta_{is}$, Eq.~\eqref{eq:6} is projected onto the corresponding eigenspace by left-multiplying with $\tilde{\mathbf{Y}}_m^\mathrm{T}$, leading to:
\begin{equation}
    \label{eq:8}
    \left(\sigma^{(p,q)}-\lambda_{m}(k)\right) \theta_{m}^{(p,q)}+f_{m}^{(p,q)}=g_{m}^{(p,q)},
\end{equation}
where
\begin{subequations}
\begin{align}
\theta_{m}^{(p,q)} &= \tilde{\mathbf{Y}}_{m}^\mathrm{T}\mathbf{B}(k)\left[\begin{array}{l}
\boldsymbol{\Psi}^{(p,q)} \\
\boldsymbol{\Upsilon}^{(p,q)}
\end{array}\right],\\
g_{m}^{(p,q)} &= \tilde{\mathbf{Y}}_{m}^\mathrm{T}\left[\begin{array}{c}
    -\mathbf{C}(k)\boldsymbol{\mu}^{(p,q)}-\mathbf{M} \boldsymbol{\nu}^{(p,q)}-\mathbf{G}^{(p,q)}-\mathbf{H}^{(p,q)} \\
    -\mathbf{M} \boldsymbol{\mu}^{(p,q)}
    \end{array}\right].
\end{align}
\end{subequations}
To mitigate the nonlinear resonances characterised by $\sigma^{(p,q)} \approx \lambda_{m}(k)$, which lead to small divisor problems, a practical approach involves eliminating $\theta_m^{(p,q)}$ via setting the suitable resonant term, $f_m^{(p,q)}$. When the resonance condition is not satisfied, the term $f_m^{(p,q)}$ is set to 0. For the resonant case, internal resonances between different dispersion solutions are excluded, as assumed at the beginning of Section~\ref{sec:3}, such that only trivial resonance relationships~\cite{TouzeCISM,RN37} are considered, namely, $\lambda_1$ and $\lambda_2$ resonate with the monomials of $\boldsymbol{\alpha}(p,q) = [n+1,n]$ and $[n,n+1]$, respectively. The CNF's resonant term is then obtained by imposing $f_{m}^{(p,q)} = g_{m}^{(p,q)}$ and $\theta_{m}^{(p,q)} = 0$ for the cases $m = 1$ and $\boldsymbol{\alpha}(p,q) = [n+1,n]$ or $m = 2$ and $\boldsymbol{\alpha}(p,q) = [n,n+1]$. By applying this parametrisation style to constrain Eq.~\eqref{eq:6} and solving it order-by-order, the nonlinear mappings $\mathbf{\Psi}(\mathbf{z}),\mathbf{\Upsilon}(\mathbf{z})$, and the corresponding reduced dynamics, can be derived. The reduced dynamics, also referred to as the CNF-BP (CNF with direct application of Bloch's assumption on physical coordinates), is expressed as
\begin{subequations}
    \label{eq:9}
    \begin{align}
    \dot{z}_1 = \lambda_{1} z_1 + \sum_{s=1}^{(o-1)/2}\varpi_sz_1^{s+1}z_2^{s},\\
    \dot{z}_2 = \bar{\lambda}_{1} z_2 + \sum_{s=1}^{(o-1)/2}\bar{\varpi}_sz_2^{s+1}z_1^{s}.
    \end{align}
\end{subequations}
By introducing the polar coordinate transformation $z_1 = \rho e^{j\psi}/2$ and $z_2 = \rho e^{-j\psi}/2$, Eqs.~\eqref{eq:9} can be rewritten as
\begin{subequations}\label{eq:10}
    \begin{align}
        \dot{\rho} &= \Re(\lambda_{1})\rho + \sum_{s = 1}\frac{\Re(\varpi_s)\rho^{2s+1}}{2^{2s}}, \label{eq:10a}\\
        \dot{\psi}&=\Im(\lambda_{1})+\sum_{s=1}\frac{\Im(\varpi_s)\rho^{2s}}{2^{2s}}, \label{eq:10b}
    \end{align}
\end{subequations}
where the operations $\Re{(\cdot)}$ and $\Im{(\cdot)}$ extract the real and imaginary parts. Eq.~\eqref{eq:10b} explicitly describes the nonlinear frequency of the metamaterial for a given wavenumber $k$. Here, we define the amplitude-dependent nonlinear frequency and damping ratio as 
\begin{subequations}
    \label{eq:add1}
    \begin{align}
    \omega_{nl} &= \dot{\psi}=\Im(\lambda_{1})+\sum_{s=1}\Im(\varpi_s)\rho^{2s}/2^{2s},\\
    \zeta_{nl} &= -\frac{\dot{\rho}}{\rho\omega_{nl}}=-\frac{\Re(\lambda_{1}) + \sum_{s = 1}\Re(\varpi_s)\rho^{2s}/2^{2s}}{\omega_{nl}}.
    \end{align}
\end{subequations} 
Notably, to enable comparison with the results obtained using EPMC-HBM~\cite{RN11} in~\ref{sec:A1}, the nonlinear damping ratio is defined with the nonlinear frequency $\omega_{nl}$ in the denominator, rather than the linear frequency. Based on both the reduced dynamics, Eqs.~\eqref{eq:10}, and the nonlinear mappings $\mathbf{\Psi}(\mathbf{z})$ and $\mathbf{\Upsilon}(\mathbf{z})$, wave attenuation properties can be effectively predicted. In the conservative case, when the losses are absent from Eqs.~\eqref{eq:2}, the eigenvalues are then purely complex conjugate, and the coefficients $\varpi_s$ are imaginary numbers~\cite{RN46}. As a consequence, one directly obtains $\dot{\rho}=0$ in Eq.~\eqref{eq:10a}, which is the consequence of the fact that in such a case, the Lyapunov subcenter manifold is densely filled by a family of periodic orbits~\cite{RN37}. Since the system is conservative, one simply retrieves that the amplitude-dependent damping $\zeta_{nl}$ is also vanishing.


As already underlined in Section~\ref{sec:1}, applying Bloch's assumption directly to the physical coordinates comes with numerous drawbacks and limitations for a nonlinear problem. First, Bloch’s assumption is inherently based on a Bloch wave formulation, expressed as $a\cos{(\omega t-kn)} = \frac{a}{2}e^{j(\omega t-kn)}+c.c.$ (where $a$ controls the wave amplitude).  However, conventional analyses often consider only the half, $\frac{a}{2}e^{j(\omega t-kn)}$, while neglecting its complex conjugate counterpart. Consequently, this approach is valid only for linear cases or nonlinear cases that do not involve nonlinear coupling between different lattice sites, such as terms of the form $\beta_1\hat{\mathbf{x}}_{\pm1}^2$. In scenarios where such coupling terms are present, the contribution of the complex conjugate component cannot be ignored, leading to potential inaccuracies in the results. This limitation implies that Eqs.~\eqref{eq:10} may yield erroneous predictions when $\mathbf{F}_{nb,2}$ and $\mathbf{F}_{nb,3}$ in Eqs.~\eqref{eq:2} are nonzero. Furthermore, Bloch’s assumption restricts the derived nonlinear dispersion equation to account accurately only for the first-order harmonic. However, in certain cases, such as those involving quadratic nonlinearities, additional wave components—specifically, the zero-order and second-order harmonics, $\frac{a\bar{a}}{2},\frac{a^2}{4}e^{2j(\omega t-kn)}$—may play a significant role. These drawbacks of the CNF-BP strategy will be further examined in Section~\ref{sec:4.1}.

\subsection{CNF-PN strategy: employing the periodic assumption on normal coordinates}
\label{sec:3.2}

To address the previously mentioned limitations of the direct application of Bloch's assumption, a second strategy to derive the CNF, is proposed. Specifically, the nonlinear change of coordinates is first applied to simplify Eqs.~\eqref{eq:2} into its normal form. Then, the periodic assumption is applied to the CNF normal coordinates rather than to the physical ones.
The CNF representation typically encapsulates the influence of higher harmonic generation within the nonlinear mappings. Given that the reduced-order model can be expressed as a form similar to Eqs.~\eqref{eq:9} with its solution represented as $z_1 = \rho e^{j\psi}/2$ and $z_2 = \rho e^{-j\psi}/2$, enforcing a periodicity assumption on the variables of the CNF is sufficient to predict dispersion characteristics and to mitigate the inadequacies of Bloch's assumption in handling non-fundamental harmonic generation. Building on this premise, the nonlinear mappings and its associated CNF are defined as $\mathbf{x}_n = \hat{\mathbf{\Psi}}(\mathbf{z}_n)$, $\mathbf{y}_n = \hat{\mathbf{\Upsilon}}(\mathbf{z}_n)$, $\dot{\mathbf{z}}_n = \hat{\mathbf{f}}(\mathbf{z}_n)$, meaning that each lattice shares the same invariant manifold, which is also aligned with the waveform invariance~\cite{Fronk2017}. The wave propagation behaviour is governed entirely by the CNF, $\dot{\mathbf{z}}_n = \hat{\mathbf{f}}(\mathbf{z}_n)$, where $\mathbf{z}_{n}$ represents the normal coordinates of $n$-th lattice, equal to $[z_{n,1};z_{n,2}]$. Based on the Bloch wave ansatz along with the inherent conjugated relationship between $z_{n,1}$ and $z_{n,2}$, the assumption $z_{n\pm1,1} = e^{\mp jk}z_{n,1}$, $z_{n\pm1,2}=e^{\pm jk}z_{n,2}$, is made. This complete periodicity assumption also resolves the previously mentioned limitation of Bloch's assumption in addressing the nonlinear coupling between adjacent lattices, particularly the effects induced by $\mathbf{F}_{nb,2}$ and $\mathbf{F}_{nb,3}$. It is also important to note that, after introducing the above periodic assumption on the normal coordinates, the subscript $n$ in $z_{n,1}$ and $z_{n,2}$ no longer plays a role in the following derivation and is thus omitted for brevity.

Using the parametrisation method and the prescribed assumptions, the following definitions are established
\begin{subequations}
    \label{eq:11}
    \begin{align}
    \mathbf{x}_n&=\hat{\mathbf{\Psi}}(\mathbf{z}_n) = \mathbf{\Phi}\mathbf{z}_n + \sum_{p=2}^{o}{\sum_{q = 1}^{q_p}{\hat{\mathbf{\Psi}}^{(p,q)}\mathbf{z}_n^{\boldsymbol{\alpha}(p,q)}}},\\
    \mathbf{x}_{n\pm1}&=\hat{\mathbf{\Psi}}(\mathbf{z}_{n\pm1}) = \mathbf{\Phi}\mathbf{E}_{\pm1}\mathbf{z}_n + \sum_{p=2}^{o}{\sum_{q = 1}^{q_p}{\hat{\mathbf{\Psi}}^{(p,q)}e^{\pm jk(\alpha_2 - \alpha_1)}\mathbf{z}_n^{\boldsymbol{\alpha}(p,q)}}},\\
    \hat{\mathbf{x}}_{\pm1} &= \mathbf{x}_n-\mathbf{x}_{n\pm1}=\mathbf{\Phi}(\mathbf{I}-\mathbf{E}_{\pm1})\mathbf{z}_n + \sum_{p = 2}^{o}\sum_{q = 1}^{q_p}\hat{\mathbf{\Psi}}_{\pm1}^{(p,q)}\mathbf{z}_n^{\boldsymbol{\alpha}(p,q)},\\
    \mathbf{y}_n & =\hat{\mathbf{\Upsilon}}(\mathbf{z}_n) = \mathbf{\Phi}\boldsymbol{\lambda}\mathbf{z}_n + \sum_{p=2}^{o}{\sum_{q=1}^{q_p}\hat{\mathbf{\Upsilon}}^{(p,q)}\mathbf{z}_n^{\boldsymbol{\alpha}(p,q)}}, \\
    \dot{\mathbf{z}}_n&=\hat{\mathbf{f}}(\mathbf{z}_n) = \boldsymbol{\lambda}\mathbf{z}_n + \sum_{p=2}^{o}\sum_{q=1}^{q_p}{\hat{\mathbf{f}}^{(p,q)}\mathbf{z}_n^{\boldsymbol{\alpha}(p,q)}},
    \end{align}
\end{subequations}
where
\begin{equation}
\mathbf{E}_{\pm1}(k) = \begin{bmatrix}
    e^{\mp jk}&0\\
    0&e^{\pm jk}
\end{bmatrix},\hat{\mathbf{\Psi}}_{\pm1}^{(p,q)} = \hat{\mathbf{\Psi}}^{(p,q)}(1-e^{\pm jk(\alpha_2-\alpha_1)}).
\end{equation}
Notably, the auxiliary index $(p,q)$ of $\alpha_1$ and $\alpha_2$ is ignored for concision where there is no ambiguity. By employing these settings, the following fundamental expressions are also introduced
\begin{subequations}
    \label{eq:12}
    \begin{align}
        \dot{\mathbf{x}}_n &= \nabla_z \hat{\mathbf{\Psi}}(\mathbf{z}_n)\hat{\mathbf{f}}(\mathbf{z}_n) = \mathbf{\Phi}\boldsymbol{\lambda}\mathbf{z}_n + \sum_{p=2}^{o}\sum_{q=1}^{q_p}\left(\sum_{s=1}^{2}\mathbf{\Phi}_s \hat{f}_s^{(p,q)} + \boldsymbol{\mu}_0^{(p,q)} + \sigma^{(p,q)}\hat{\mathbf{\Psi}}^{(p,q)}\right)\mathbf{z}_n^{\boldsymbol{\alpha}(p,q)},\\
        \dot{\mathbf{x}}_{n\pm1} &= \nabla_z \hat{\mathbf{\Psi}}(\mathbf{E}_{\pm1}\mathbf{z}_n)\hat{\mathbf{f}}(\mathbf{z}_n) = \mathbf{\Phi}\mathbf{E}_{\pm1}\boldsymbol{\lambda}\mathbf{z}_n + \\
        &\sum_{p=2}^{o}\sum_{q=1}^{q_p}\left(\sum_{s=1}^{2}\mathbf{\Phi}_s \mathbf{E}_{\pm1s}\hat{f}_s^{(p,q)} + \boldsymbol{\mu}_{\pm1}^{(p,q)} + \sigma^{(p,q)}e^{\pm jk(\alpha_2 - \alpha_1)}\hat{\mathbf{\Psi}}^{(p,q)}\right)\mathbf{z}_n^{\boldsymbol{\alpha}(p,q)},\\
        \dot{\mathbf{y}}_n &= \nabla_z \hat{\mathbf{\Upsilon}}(\mathbf{z}_n)\hat{\mathbf{f}}(\mathbf{z}_n) = \mathbf{\Phi}\boldsymbol{\lambda}^2\mathbf{z}_n + \sum_{p=2}^{o}\sum_{q=1}^{q_p}\left(\sum_{s=1}^{2}\mathbf{\Phi}_s \lambda_s \hat{f}_s^{(p,q)} + \boldsymbol{\nu}_0^{(p,q)} + \sigma^{(p,q)}\hat{\mathbf{\Upsilon}}^{(p,q)}\right)\mathbf{z}_n^{\boldsymbol{\alpha}(p,q)},\\
        \mathbf{F}_{n,2} &= \sum_{p=2}^{o}\sum_{q=1}^{q_p} \mathbf{G}_0^{(p,q)}\mathbf{z}_n^{\boldsymbol{\alpha}(p,q)},\mathbf{F}_{nb,2} = \sum_{p=2}^{o}\sum_{q=1}^{q_p} \mathbf{G}_{\pm1}^{(p,q)}\mathbf{z}_n^{\boldsymbol{\alpha}(p,q)},\\
        \mathbf{F}_{n,3} &= \sum_{p=3}^{o}\sum_{q=1}^{q_p} \mathbf{H}_0^{(p,q)}\mathbf{z}_n^{\boldsymbol{\alpha}(p,q)},\mathbf{F}_{nb,3} = \sum_{p=3}^{o}\sum_{q=1}^{q_p} \mathbf{H}_{\pm1}^{(p,q)}\mathbf{z}_n^{\boldsymbol{\alpha}(p,q)},
    \end{align}
\end{subequations}
where
\begin{subequations}
\label{eq:22a-e}
\begin{align}
\boldsymbol{\mu}_{\pm1}^{(p,q)} &= \sum_{s=1}^{2}\sum_{p_1=2}^{p-1}\sum_{q_1=1}^{q_{p1}}\alpha_s(p-p_1+1,q_2)\hat{\mathbf{\Psi}}^{(p-p_1+1,q_2)}e^{\pm jk(\alpha_2(p-p_1+1,q_2) - \alpha_1(p-p_1+1,q_2))}\hat{f}_s^{(p_1,q_1)},\\
\mathbf{G}_0^{(p,q)} &= \sum_{p_1=1}^{p-1}\sum_{q_1=1}^{q_{p1}}\mathbf{F}_{n,2}(\hat{\mathbf{\Psi}}^{(p_1,q_1)},\hat{\mathbf{\Psi}}^{(p-p_1,q_2)}),\\
\mathbf{G}_{\pm1}^{(p,q)} &= \sum_{p_1=1}^{p-1}\sum_{q_1=1}^{q_{p1}}\mathbf{F}_{nb,2}(\hat{\mathbf{\Psi}}_{\pm1}^{(p_1,q_1)},\hat{\mathbf{\Psi}}_{\pm1}^{(p-p_1,q_2)}),\\
\mathbf{H}_0^{(p,q)} &= \sum_{p_1=1}^{p-2}\sum_{p_2=1}^{p-p_1-1}\sum_{q_1=1}^{q_{p1}}\sum_{q_2=1}^{q_{p2}}\mathbf{F}_{n,3}(\hat{\mathbf{\Psi}}^{(p_1,q_1)},\hat{\mathbf{\Psi}}^{(p_2,q_2)},\hat{\mathbf{\Psi}}^{(p-p_1-p_2,m_3)}),\\
\mathbf{H}_{\pm1}^{(p,q)} &= \sum_{p_1=1}^{p-2}\sum_{p_2=1}^{p-p_1-1}\sum_{q_1=1}^{q_{p1}}\sum_{q_2=1}^{q_{p2}}\mathbf{F}_{nb,3}(\hat{\mathbf{\Psi}}_{\pm1}^{(p_1,q_1)},\hat{\mathbf{\Psi}}_{\pm1}^{(p_2,q_2)},\hat{\mathbf{\Psi}}_{\pm1}^{(p-p_1-p_2,m_3)}).
\end{align}
\end{subequations}
Note that, in Eqs.~\eqref{eq:22a-e}, and in order not to introduce too many symbols, $\boldsymbol{\mu}_0^{(p,q)}$ and $\boldsymbol{\nu}_0^{(p,q)}$ are simply taken equal to $\boldsymbol{\mu}^{(p,q)}$ and $\boldsymbol{\nu}^{(p,q)}$ in Eq.~\eqref{eq:7}. The $\mathbf{E}_{\pm1s}$ and $\hat{f}_s^{(p,q)}$ denote the $s$-th elements in $\mathrm{diag}(\mathbf{E}_{\pm1})$ and $\hat{\mathbf{f}}^{(p,q)}$. The remaining variables in Eqs.~\eqref{eq:12} are defined in the same manner as in Eq.~\eqref{eq:7}. Note that these expressions of nonlinear forces in Eqs.~\eqref{eq:12} are straightforward, without the need for further explanation, whereas a more detailed derivation of $\dot{\mathbf{x}}_n,\dot{\mathbf{x}}_{n\pm1}$, and $\dot{\mathbf{y}}_n$ is provided in~\ref{sec:Add}. Substituting Eqs.~\eqref{eq:11} and~\eqref{eq:12} into Eqs.~\eqref{eq:2} yields the invariance equation:
\begin{subequations}
\label{eq:13}
\begin{align}
    \begin{aligned}
        \left[\mathbf{M}\mathbf{\Phi}\boldsymbol{\lambda}^2 + (\mathbf{C}_0\mathbf{\Phi}+\mathbf{C}_{-1}\mathbf{\Phi}(\mathbf{E}_{-1}+\mathbf{E}_{+1}))\boldsymbol{\lambda}+(\mathbf{K}_0\mathbf{\Phi}+\mathbf{K}_{-1}\mathbf{\Phi}(\mathbf{E}_{-1}+\mathbf{E}_{+1}))\right]\mathbf{z}_n + \\
        \sum_{p=2}^{o}\sum_{q=1}^{q_p} \left[\sigma^{(p,q)}(\mathbf{M}\hat{\mathbf{\Upsilon}}^{(p,q)}+(\mathbf{C}_0+2\cos(\alpha_2-\alpha_1)k\mathbf{C}_{-1})\hat{\mathbf{\Psi}}^{(p,q)})+(\mathbf{K}_0+2\cos{(\alpha_2-\alpha_1)k}\mathbf{K}_{-1})\hat{\mathbf{\Psi}}^{(p,q)}\right.\\
        \left.\sum_{s=1}^{2}\hat{f}_s^{(p,q)}(\lambda_s\mathbf{M}+(\mathbf{C}_0+\mathbf{C}_{-1}(\mathbf{E}_{-1}+\mathbf{E}_{+1})))\mathbf{\Phi}_s + \mathbf{M}\boldsymbol{\nu}_0^{(p,q)}+\sum_{i=-1}^{1}\mathbf{C}_i\boldsymbol{\mu}_i^{(p,q)}+\mathbf{G}_{i}^{(p,q)}+\mathbf{H}_i^{(p,q)}\right]\mathbf{z}_n^{\boldsymbol{\alpha}(p,q)}=0,
        \end{aligned}\\
        (\mathbf{M}\mathbf{\Phi}\boldsymbol{\lambda}-\mathbf{M}\mathbf{\Phi}\boldsymbol{\lambda})\mathbf{z}_n + \sum_{p=2}^{o}\sum_{q=1}^{q_p}\left[\sigma^{(p,q)}\mathbf{M}\hat{\mathbf{\Psi}}^{(p,q)}-\mathbf{M}\hat{\mathbf{\Upsilon}}^{(p,q)}+\sum_{s=1}^{2}\hat{f}_s^{(p,q)}\mathbf{M}\mathbf{\Phi}_s+\mathbf{M}\boldsymbol{\mu}_0^{(p,q)}\right]=0.
    \end{align}
\end{subequations}

Using the relationship: $\mathbf{E}_{-1}+\mathbf{E}_{+1} = 2\cos{k}\mathbf{I}$, where $\mathbf{I}$ denotes the identity matrix, the first-order problem, which corresponds to the linear solution, and the arbitrary $p$-th order problem, can be obtained. Notably, the first-order linear problem with the present CNF-PN strategy is identical to that of the CNF-BP strategy. Indeed, the parametrisation method as well as the normal form approach assumes identity-tangency to the linear solutions, such that in both cases the linear problem is left unchanged, only the treatment of the nonlinear terms is modified. The first-order problem is thus automatically satisfied due to the application of linear dispersion solutions with $\mathbf{\Phi}$ and $\boldsymbol{\lambda}$.

Regarding the order-$p$ problem, it can be expressed in the CNF-PN method as:
\begin{equation}
    \label{eq:14}
    \begin{aligned}
        \left(\sigma^{(p,q)}\mathbf{B}((\alpha_2-\alpha_1)k) - \mathbf{A}((\alpha_2-\alpha_1)k)\right)\begin{bmatrix}
            \hat{\mathbf{\Psi}}^{(p,q)}\\
            \hat{\mathbf{\Upsilon}}^{(p,q)}
        \end{bmatrix}
        +\sum_{s=1}^{2}\hat{f}_s^{(p,q)}\mathbf{B}(k)\begin{bmatrix}
                \mathbf{\Phi}_s(k)\\
                \mathbf{\Phi}_s(k)\lambda_s(k)
            \end{bmatrix}\\=\begin{bmatrix}
                -\sum_{i=-1}^1\mathbf{C}_i\boldsymbol{\mu}_{i}^{(p,q)}-\sum_{i=-1}^1\mathbf{G}_i^{(p,q)}-\sum_{i=-1}^1\mathbf{H}_i^{(p,q)}-\mathbf{M}\boldsymbol{\nu}_0^{(p,q)}\\
                -\mathbf{M}\boldsymbol{\mu}_0^{(p,q)}
            \end{bmatrix},
    \end{aligned}
\end{equation}
where $\sigma^{(p,q)} = \alpha_1\lambda_1 + \alpha_2\lambda_2$. Note that, in the presented derivation and in order to make the presentation easier, it is assumed that the computation of the acoustic nonlinear dispersion relationship is searched for. Therefore the selected eigenvalues are $\lambda_{1,2} = -\zeta_a \omega_a \pm j\omega_a \sqrt{1-\zeta_a^2}$. Of course the same derivation holds for the optic branch, one has then just to replace $(\lambda_1,\lambda_2) = (\Lambda_1,\Lambda_2)$ by $(\lambda_1,\lambda_2) =(\Lambda_3,\Lambda_4)$ and modify the subscripts $a$ (for acoustic) in the eigenvalues by $o$, see Section~\ref{sec:2.2}.

Using the known expression of the conjugate eigenvalues, one can rewrite the combination $\sigma^{(p,q)}$ as:
\begin{equation}
    \sigma^{(p,q)} = -(\alpha_1 + \alpha_2)\zeta_{a}\omega_{a} + n_{(p,q)} j \omega_{a}\sqrt{1-\zeta_{a}^2},
\end{equation}
where the notation $n_{(p,q)}= \alpha_1 - \alpha_2$ has been introduced. This is indeed meaningful since one can observe that the matrices $\mathbf{B}$ and $\mathbf{A}$ in the first term of Eq.~\eqref{eq:14} are not computed at the value $k$ of the wavenumber, but instead at $ n_{(p,q)} k$. This fact will have very important consequences on the resonance relationship that is key in the solving process of the DPIM.
Projecting Eq.~\eqref{eq:14} onto the associated dispersion branch's eigenspace is a necessary step to clearly make appear the solutions at the level of each monomial. Since the linear operators are computed at the combination wavenumber value $-n_{(p,q)} k=(\alpha_2 - \alpha_1)k$, the projection step is operated here through a left multiplication by the vector $\tilde{\mathbf{Y}}_{m}(n_{(p,q)}k)^\mathrm{T}$ ($\mathbf{Y}_{m}(n_{(p,q)}k)^\mathrm{T}$ for acoustic calculation, $\mathbf{Y}_{m+2}(n_{(p,q)}k)^\mathrm{T}$ for optic case), which is equivalent to $\tilde{\mathbf{Y}}_{m}(-n_{(p,q)}k)^\mathrm{T}$ due to the symmetry of $\mathbf{A}(n_{(p,q)}k)$ and $\mathbf{B}(n_{(p,q)}k)$ as a function of $n_{(p,q)}k$. After this operation, the following expression is obtained~:
\begin{equation}
    \label{eq:15}
    \left(\sigma^{(p,q)}-\lambda_m(n_{(p,q)}k)\right)\theta_m^{(p,q)} + \sum_{s=1}^{2}\hat{f}_s^{(p,q)}\tilde{\mathbf{Y}}_{m}(n_{(p,q)}k)^\mathrm{T}\mathbf{B}(k)\tilde{\mathbf{Y}}_{s}(k) = g_m^{(p,q)},
\end{equation}
where
\begin{subequations}
\begin{align}
\theta_m^{(p,q)} &= \tilde{\mathbf{Y}}_{m}(n_{(p,q)}k)^\mathrm{T}\mathbf{B}(n_{(p,q)}k)\begin{bmatrix}\hat{\mathbf{\Psi}}^{(p,q)}\\
\hat{\mathbf{\Upsilon}}^{(p,q)}\end{bmatrix},\\
g_m^{(p,q)} &= \tilde{\mathbf{Y}}_{m}(n_{(p,q)}k)^\mathrm{T}\begin{bmatrix}
                -\sum_{i=-1}^1\mathbf{C}_i\boldsymbol{\mu}_{i}^{(p,q)}-\sum_{i=-1}^1\mathbf{G}_i^{(p,q)}-\sum_{i=-1}^1\mathbf{H}_i^{(p,q)}-\mathbf{M}\boldsymbol{\nu}_0^{(p,q)}\\
                -\mathbf{M}\boldsymbol{\mu}_0^{(p,q)}
            \end{bmatrix}.
\end{align}
\end{subequations}

Notably, in the second term of Eq.\eqref{eq:14}, the matrix $\mathbf{B}(k)$ is the function of $k$ rather than $n_{(p,q)}k$, causing that the second term in Eq.~\eqref{eq:15} cannot be simplified via the normalisation condition of eigenvector. Turning attention to the first term in Eq.\eqref{eq:15}, it makes now clearly appear the nonlinear resonance relationships, that are notably different from those derived with the CNF-BP strategy, see Eq.~\eqref{eq:8}, the distinctive feature being that the combination of master eigenfrequencies appearing in $\sigma^{(p,q)}$ now resonates with an eigenvalue $\lambda_m$ computed for the wavenumber $n_{(p,q)}k$ rather than $k$. 
 In line with the assumption of light damping~\cite{TouzeCISM,RN37}, the resonance relationship $\sigma^{(p,q)} \simeq \lambda_m(n_{(p,q)}k)$,  creates small divisors and needs to be avoided. Taking acoustic calculation as an example ($(\lambda_1,\lambda_2)=(\Lambda_1,\Lambda_2)$), it can be checked only on the oscillatory (imaginary) part, neglecting the real parts, yielding:
\begin{equation}\label{eq:resonanceCNFPN}
    n_{(p,q)}\omega_{a}(k)\sqrt{1-\zeta_{a}^2} \simeq \Im(\lambda_m(n_{(p,q)}k)),
\end{equation}
where $\Im$ stands for the imaginary part. Notably, the resonance condition given in Eq.~\eqref{eq:resonanceCNFPN} is equivalent to the internal wave–wave resonance condition discussed in~\cite{RN23}. However, the present work excludes internal resonances between different dispersion solutions. As a result, only two specific scenarios consistently lead to the emergence of small divisors in Eq.~\eqref{eq:15}. 
The first is the trivial resonance relationships, where $n_{(p,q)}= 1$ or -1. This implies that $\lambda_1$ and $\lambda_2$ may generate resonant terms of $\boldsymbol{\alpha}(p,q) = [n+1,n]$ or $[n,n+1]$, $n = 1,2,\cdots$, a phenomenon also observed in the direct application of Bloch's assumption. The second scenario occurs in the acoustic branch, which is not accounted for in the BP strategy. Here, when $n_{(p,q)} = 0$, the vanishing eigenvalues $\lambda_1(0) = \lambda_2(0) = \Lambda_1(0) = \Lambda_2(0) = 0$, as highlighted in Section~\ref{sec:2.2}, may resonate with terms of $\boldsymbol{\alpha}(p,q) = [n, n]$. This corresponds to the case of $0\times\Im{(\lambda_1(k))}=\Im(\lambda_m(0\times k))$. 

Note that if the nonlinearities between the host mass blocks are ignored, and the focus is restricted to the case of $\boldsymbol{\alpha}(p,q) = [n+1,n]$ or $[n,n+1]$, Eq.~\eqref{eq:15} reduces to the same form as Eq.~\eqref{eq:8}, which is based on the direct application of Bloch's assumption. In fact, the terms, $z_{1}^{n+1}z_{2}^n$ and $z_{1}^nz_{2}^{n+1}$, do not generate non-fundamental harmonics. This implies that directly applying Bloch's assumption and imposing periodic assumptions to the normal coordinates will yield a similar CNF and invariant manifold when the nonlinear wave's energy is predominantly concentrated in the fundamental harmonic.

Let us now analyse the resonance conditions and consequences on the CNF-PN derivation in more detail.

\subsubsection{Case of $\boldsymbol{\alpha}(p,q) = [n+1,n]$ or $[n, n+1]$}
\label{sec:3.2.1}

This case corresponds to trivial resonances that are commonly encountered in vibration theory~\cite{TouzeCISM,RN37}, and appears here in the particular case of dispersion spectrum when $\boldsymbol{\alpha}(p,q) = [n+1,n]$ or $[n, n+1]$ and $n_{(p,q)} = \pm1$. To draw out simple expressions for the nonliear dispersion curves, the normal form style is selected, and Eq.~\eqref{eq:15} is solved with: $\theta_m^{(p,q)} = 0,f_m^{(p,q)} = g_m^{(p,q)}$ for $m=1$ and $n_{(p,q)} = 1$, or $m=2$ and $n_{(p,q)}=-1$. Under this setting, Eq.~\eqref{eq:14} can be reformulated into a well-posed form and solved in a direct manner using the bordering technique proposed in~\cite{Carini2015,RN43,Vizzaccaro2024}, which implies rewriting the condition $\theta_m^{(p,q)} = 0$ as:
\begin{equation}\label{eq:ziborder}
    \mathbf{\Phi}_{m}^\mathrm{T}(\mathbf{C}(k) +\lambda_m\mathbf{M})\hat{\mathbf{\Psi}}^{(p,q)}+\mathbf{\Phi}_{m}^\mathrm{T}\mathbf{M}\hat{\mathbf{\Upsilon}}^{(p,q)}=\mathbf{0}.    
\end{equation}
The bordering technique consists in rewriting Eq.~\eqref{eq:14} in a solvable manner by bordering the singular part 
by the right and left eigenvectors of its kernel. This is realised using Eq.~\eqref{eq:ziborder}, yielding:
\begin{equation}
    \label{eq:16}
    \begin{aligned}
    \begin{bmatrix}
        \mathbf{K}(k) +\sigma^{(p,q)}\mathbf{C}(k) & \sigma^{(p,q)}\mathbf{M}&(\mathbf{C}(k) +\lambda_m\mathbf{M})\mathbf{\Phi}_m\\
        \sigma^{(p,q)}\mathbf{M} &  -\mathbf{M}& \mathbf{M}\mathbf{\Phi}_m\\
        \mathbf{\Phi}_{m}^\mathrm{T}(\mathbf{C}(k) +\lambda_m\mathbf{M}) &\mathbf{\Phi}_{m}^{\mathrm{T}}\mathbf{M}&\mathbf{0}
    \end{bmatrix}\begin{bmatrix}
        \hat{\mathbf{\Psi}}^{(p,q)}\\
        \hat{\mathbf{\Upsilon}}^{(p,q)}\\
        \hat{f}_m^{(p,q)}
    \end{bmatrix}=\\
    \begin{bmatrix}
                -\sum_{i=-1}^1\mathbf{C}_i\boldsymbol{\mu}_{i}^{(p,q)}-\sum_{i=-1}^1\mathbf{G}_i^{(p,q)}-\sum_{i=-1}^1\mathbf{H}_i^{(p,q)}-\mathbf{M}\boldsymbol{\nu}_0^{(p,q)}\\
                -\mathbf{M}\boldsymbol{\mu}_0^{(p,q)}\\
                \mathbf{0}
            \end{bmatrix}.
            \end{aligned}
\end{equation}
Note that a simpler problem can still be derived, by eliminating the terms of the velocity mapping, and express them as functions of the displacement mapping only, hence dividing by two the size of the problems to be solved~\cite{RN43}. From the second equation in Eq.~\eqref{eq:16}, the following relationship is obtained
\begin{equation}
\label{eq:17}
    \hat{\mathbf{\Upsilon}}^{(p,q)} = \sigma^{(p,q)}\hat{\mathbf{\Psi}}^{(p,q)} + \mathbf{\Phi}_m\hat{f}_m^{(p,q)} + \boldsymbol{\mu}_0^{(p,q)}.
\end{equation}
Substituting Eq.~\eqref{eq:17} into Eq.~\eqref{eq:16} leads to
\begin{equation}
    \label{eq:18}
    \begin{bmatrix}
        \sigma^{(p,q)2}\mathbf{M}+\sigma^{(p,q)}\mathbf{C}(k) + \mathbf{K}(k)&\left((\sigma^{(p,q)}+\lambda_m)\mathbf{M}+\mathbf{C}(k)\right)\mathbf{\Phi}_m\\
        \mathbf{\Phi}_{m}^\mathrm{T}(\mathbf{C}(k)+(\sigma^{(p,q)}+\lambda_m)\mathbf{M}) &\mathbf{\Phi}_{m}^{\mathrm{T}}\mathbf{M}\mathbf{\Phi}_m
    \end{bmatrix}\begin{bmatrix}
        \hat{\mathbf{\Psi}}^{(p,q)}\\
        \hat{f}_m^{(p,q)}
    \end{bmatrix}=
    \begin{bmatrix}
        \mathbf{\Xi}^{(p,q)}\\
        -\mathbf{\Phi}_{m}^{\mathrm{T}}\mathbf{M}\boldsymbol{\mu}_0^{(p,q)}
    \end{bmatrix},
\end{equation}
where
\begin{equation}
\mathbf{\Xi}^{(p,q)} = -\sum_{i=-1}^1\mathbf{C}_i\boldsymbol{\mu}_{i}^{(p,q)}-\sum_{i=-1}^1\mathbf{G}_i^{(p,q)}-\sum_{i=-1}^1\mathbf{H}_i^{(p,q)}-\mathbf{M}\boldsymbol{\nu}_0^{(p,q)}-\sigma^{(p,q)}\mathbf{M}\boldsymbol{\mu}_0^{(p,q)}.
\end{equation}

Thanks to Eq.~\eqref{eq:18}, a direct solution is finally derived for the unknowns $\hat{\mathbf{\Psi}}^{(p,q)}$ and $\hat{f}_m^{(p,q)}$, such that arbitrary order solutions in this CNF-PN strategy are automatically derived. Let us now detail the last particular case of resonance that arises from the nonlinear resonance relationships, Eq.~\eqref{eq:resonanceCNFPN}.

\subsubsection{Case of $\boldsymbol{\alpha}(p,q) = [n,n]$ in acoustic branch}
\label{sec:3.2.2}
For the acoustic case of $\boldsymbol{\alpha}(p,q) = [n,n]$, or equivalently when $n_{(p,q)} = 0$, the eigenvalues $\lambda_1 = \lambda_2 = 0$ resonate with the terms associated with $\boldsymbol{\alpha}(p,q) = [n,n]$. However, since both eigenvalues share the same eigenvector, as highlighted in Section~\ref{sec:2.2}, the standard parametrisation approach based on the CNF style becomes inapplicable.

For providing novel constraint equations, it is important to build the relationship between $\hat{\mathbf{\Upsilon}}^{(p,q)}$ and $\hat{\mathbf{\Psi}}^{(p,q)}$ via the second equation in Eq.~\eqref{eq:14}, given by
\begin{equation}
\label{eq:20}
    \hat{\mathbf{\Upsilon}}^{(p,q)} = \sigma^{(p,q)}\hat{\mathbf{\Psi}}^{(p,q)}+\mathbf{\Phi}_1\hat{f}_1^{(p,q)}+\mathbf{\Phi}_2\hat{f}_2^{(p,q)}+\boldsymbol{\mu}_0^{(p,q)}.
\end{equation}
Exploiting it, Eq.~\eqref{eq:14} can be simplified as 
\begin{equation}
\label{eq:ind}
\begin{aligned}
    &(\mathbf{K}(0)+\sigma^{(p,q)}\mathbf{C}(0)+\sigma^{(p,q)2}\mathbf{M})\hat{\mathbf{\Psi}}^{(p,q)}+(\mathbf{C}(k)+(\lambda_1+\sigma^{(p,q)})\mathbf{M})\mathbf{\Phi}_1\hat{f}_1^{(p,q)} \\&+ (\mathbf{C}(k)+(\lambda_2+\sigma^{(p,q)})\mathbf{M})\mathbf{\Phi}_2\hat{f}_2^{(p,q)} = \mathbf{\Xi}^{(p,q)}.
    \end{aligned}
\end{equation}
Notably, for $\boldsymbol{\alpha}(p,q) = [n,n]$, both $\hat{\mathbf{\Psi}}^{(p,q)},\sigma^{(p,q)}$, and $\mathbf{\Xi}^{(p,q)}$ are real, and the terms $(\mathbf{C}(k)+(\lambda_1+\sigma^{(p,q)})\mathbf{M})\mathbf{\Phi}_1$ and $(\mathbf{C}(k)+(\lambda_2+\sigma^{(p,q)})\mathbf{M})\mathbf{\Phi}_2$ form a conjugate pair. Consequently, to ensure that the expression $(\mathbf{C}(k)+(\lambda_1+\sigma^{(p,q)})\mathbf{M})\mathbf{\Phi}_1\hat{f}_1^{(p,q)}+(\mathbf{C}(k)+(\lambda_2+\sigma^{(p,q)})\mathbf{M})\mathbf{\Phi}_2\hat{f}_2^{(p,q)}$ yields real-valued results for the solvability of Eq.~\eqref{eq:ind}, $\hat{f}_1^{(p,q)}$ and $\hat{f}_2^{(p,q)}$ must be complex conjugates. To enforce this condition and eliminate the singularity from $\mathbf{K}(0)$ and $\mathbf{C}(0)$, we introduce the following constraints
\begin{equation}
\label{eq:19}
    \mathbf{\Phi}_0^\mathrm{T}\hat{\mathbf{\Psi}}^{(p,q)} = 0,\hat{f}_1^{(p,q)} = -\hat{f}_2^{(p,q)},\mathbf{\Phi}_0=[1;0].
\end{equation}
Using the above settings, Eq.~\eqref{eq:ind} can be rewritten in a solvable manner as:
\begin{equation}
\label{eq:21}
    \begin{aligned}
    \begin{bmatrix}
        \mathbf{K}(0) + \sigma^{(p,q)}\mathbf{C}(0) +\sigma^{(p,q)2}\mathbf{M}& (\mathbf{C}(k)+(\lambda_1+\sigma^{(p,q)})\mathbf{M})\mathbf{\Phi}_1&(\mathbf{C}(k)+(\lambda_2+\sigma^{(p,q)})\mathbf{M})\mathbf{\Phi}_2\\
        0 &1&1\\
        \mathbf{\Phi}_0^\mathrm{T}&\mathbf{0}&\mathbf{0}
    \end{bmatrix}\begin{bmatrix}
        \hat{\mathbf{\Psi}}^{(p,q)}\\
        \hat{f}_1^{(p,q)}\\
        \hat{f}_2^{(p,q)}
    \end{bmatrix}=\\
    \begin{bmatrix}
                \mathbf{\Xi}^{(p,q)}\\
                0\\
                0
            \end{bmatrix}.
            \end{aligned}
\end{equation}

\subsubsection{Case of non-resonant term}
\label{sec:3.2.3}
For non-resonant terms, the coefficients of the invariant manifold and CNF-PN can be obtained by solving
\begin{subequations}
\label{eq:22}
    \begin{align}
        \hat{\mathbf{\Psi}}^{(p,q)} = (\sigma^{(p,q)2}\mathbf{M}+\sigma^{(p,q)}\mathbf{C}(n_{(p,q)}k)+\mathbf{K}(n_{(p,q)}k))^{-1}\mathbf{\Xi}^{(p,q)},\\
        \hat{\mathbf{\Upsilon}}^{(p,q)} = \sigma^{(p,q)}\hat{\mathbf{\Psi}}^{(p,q)}+\boldsymbol{\mu}_0^{(p,q)},\\
        \hat{f}_1^{(p,q)} = \hat{f}_2^{(p,q)} = 0.
    \end{align}
\end{subequations}
\subsubsection{Complex Normal form and invariant manifold}
Using the above setting, the final reduced-order model, also referred to as CNF-PN, and the corresponding invariant manifold can be obtained. The resulting CNF-PN differs slightly from Eqs.~\eqref{eq:9} and takes the following form:
\begin{subequations}
    \label{eq:23}
    \begin{align}
    \dot{z}_1 = \lambda_{1} z_1 + \sum_{s=1}^{(o-1)/2}\varpi_sz_1^{s+1}z_2^{s}+\sum_{n=1}^{o/2}\varpi_{0,n}jz_1^nz_2^n,\\
    \dot{z}_2 = \bar{\lambda}_{1} z_2 + \sum_{s=1}^{(o-1)/2}\bar{\varpi}_sz_2^{s+1}z_1^{s}-\sum_{n=1}^{o/2}\varpi_{0,n}jz_1^nz_2^n.
    \end{align}
\end{subequations}
Setting $z_1 = \rho e^{j\psi}/2$ and $z_2 = \rho e^{-j\psi}/2$, Eqs.~\eqref{eq:23} can be rewritten as
\begin{subequations}
    \label{eq:24}
    \begin{align}
        \dot{\rho} = \Re(\lambda_{1})\rho + \sum_{s = 1}^{o/2-1}\frac{\Re(\varpi_s)\rho^{2s+1}}{2^{2s}}+\sum_{n=1}^{o/2}\frac{\varpi_{0,n}\rho^{2n}\sin{\psi}}{2^{2n-1}},\\
        \dot{\psi}=\Im(\lambda_{1})+\sum_{s=1}^{o/2}\frac{\Im(\varpi_s)\rho^{2s}}{2^{2s}}+\sum_{n=1}^{o/2}\frac{\varpi_{0,n}\rho^{2n-1}\cos{\psi}}{2^{2n-1}}.
    \end{align}
\end{subequations}
Notably, the last terms, $\varpi_{0,n}\rho^{2n}\sin{\psi}/2^{2n-1}, \varpi_{0,n}\rho^{2n-1}\cos{\psi}/2^{2n-1}$, in the above Eqs.~\eqref{eq:24} are zero for the optic branch or periodic for the acoustic branch. These terms arise specifically from the resonant terms of $\boldsymbol{\alpha}(p,q) = [n,n]$ in the CNF-PN, and therefore do not appear in CNF-BP. However, since these terms are periodic over $\psi \in [0, 2\pi]$, their effect is eliminated upon averaging, and they can thus be safely neglected. The remaining variables retain the same symbols and meanings as in CNF-BP, Eq.~\eqref{eq:9}, although their values for $\varpi_s$ may differ due to the distinct method used to impose the periodic assumption. 
Following these simplifications, the damping ratio and nonlinear frequency are the same as Eqs.~\eqref{eq:add1}. 


\section{Numerical results}
\label{sec:4}

Section~\ref{sec:3} introduced two strategies for handling shared variables in an infinite periodic chain and deriving the corresponding complex normal forms. In this Section, their respective advantages are further discussed using numerical simulations and extensive comparisons. A numerical solution for the nonlinear dispersion curves is obtained thanks to the EPMC-HBM procedure shown in~\cite{RN11}, and taken as a reference. The predictions given by both CNF-BP and CNF-PN are first contrasted for two distinct undamped nonlinear periodic chains. This comparison highlights the shortcomings of the CNF-BP and the advancements introduced by the CNF-PN strategy. Specifically, two cases are examined: one with quadratic nonlinearity $\beta_2$ in the attachments, illustrating the CNF-BP’s inaccuracy in capturing the generation of zero-order and higher-order harmonics, and another with cubic nonlinearity $\gamma_1$ in the host oscillators, demonstrating the CNF-BP’s inability to account for nonlinearity in the host mass. Notably, when the only nonlinearity present is the cubic stiffness $\gamma_2$ in the attachments, the BP strategy typically yields accurate results, as extensively validated in the literature~\cite{RN16,RN27,RN28}. Using the proposed computational framework, it is straightforward to demonstrate that CNF-BP and CNF-PN with $o=3$ and only $\gamma_2 \neq 0$ for the nonlinearities, provide the same expression for the reduced dynamics, which reads:
\begin{equation}
z_1 = \lambda_{1}z_1-3\gamma_2\mathbf{\Phi}_{21}^3\bar{\mathbf{\Phi}}_{21}z_1^2z_2.
\end{equation}
where $\mathbf{\Phi}_{is}$ denotes the element of the $i$-th row and $s$-th column in matrix $\mathbf{\Phi}$, which is equal to $[\mathbf{\phi}_1,\mathbf{\phi}_2]$ for acoustic calculation or $[\mathbf{\phi}_3,\mathbf{\phi}_4]$ for optic case. 
 Therefore, for brevity, no corresponding example is provided in Section~\ref{sec:4.1}. After confirming the advantages of the PN strategy, Section~\ref{sec:4.2} investigates the effect of the truncation order $o$ in the CNF-PN using an undamped chain with cubic nonlinearity $\gamma_2$. Subsequently, Section~\ref{sec:4.3} incorporates damping terms, $c_1$ and $c_2$, and examines two types of nonlinear metamaterial chains—one possessing only cubic nonlinearity $\gamma_2$ and another with multiple nonlinearities, including $\beta_1, \beta_2, \gamma_1$, and $\gamma_2$. The CNF-PN is then compared against two reference methods, EPMC-HBM~\cite{RN11} and MMS~\cite{RN30,RN31,RN32}, further highlighting its advantages. Additionally, a finite annular chain comprising 100 lattices is analysed to provide numerical integration results and illustrate the wave attenuation process. These results serve as benchmarks to assess the accuracy and performance of the three methods. 
Regarding the two reference methods, their detailed introductions are provided in~\ref{sec:A1} (for EPMC-HBM) and~\ref{sec:A2} (for MMS).

\subsection{Limitations of the CNF-BP Strategy}
\label{sec:4.1}
As discussed in Sections~\ref{sec:1} and~\ref{sec:3.1}, the direct application of Bloch’s assumption in physical coordinates presents two primary limitations: (i) inaccurate predictions of zero-order and high-order harmonics and (ii) an inability to account for nonlinearities in host oscillators. To illustrate these shortcomings, two conservative cases are first examined: a nonlinear chain incorporating quadratic nonlinearity, $\beta_2$, in the attachments and a chain exhibiting cubic nonlinearity, $\gamma_1$, in the host masses. The parameter values for all subsequent numerical examples are provided in Table~\ref{tab:1}. To facilitate an analytical comparison, the expressions for the coefficient $\varpi_1$ in CNF-BP and CNF-PN are also presented in Table~\ref{tab:2}, highlighting the differences in terms involving the wavenumber $k$. 

\begin{table}[h!]
  \centering
  \caption{Parameters of the infinite periodic chain.}
  \label{tab:1}       
  \begin{tabular}{ll}
  \hline\noalign{\smallskip}
  Parameters & Values \\
  \noalign{\smallskip}\hline\noalign{\smallskip}
  Mass of host oscillator, $m_{h}$ & 1 ($\mathrm{kg}$)\\
  Mass of attachment, $m_{a}$ & 0.5 ($\mathrm{kg}$)\\
  Linear stiffness between host oscillator, $\kappa_1$ & 1 ($\mathrm{N/m}$)\\
  Linear stiffness in attachment, $\kappa_2$ & 0.4 ($\mathrm{N/m}$)\\
  \noalign{\smallskip}\hline
  \end{tabular}
\end{table}

\begin{table}[h!]
  \centering
  \caption{The coefficients, $\varpi_1$, of CNF-BP and CNF-PN for nonlinear metamaterial chain with $\beta_2$ or $\gamma_1$. Using the parameters in Table~\ref{tab:1} and setting $c_1=c_2=0$. Note: the acoustic and optic CNF-PNs share the same $\varpi_1$ in these examples. Here, only the acoustic case is presented; the optic CNF-PNs can be obtained by simply replacing the linear frequency's subscript $a$ with $o$.}
  \label{tab:2}       
  \begin{tabular}{ll}
  \hline\noalign{\smallskip}
  Normal form & Leading order nonlinear coefficient $\varpi_1$ \\
  \noalign{\smallskip}\hline\noalign{\smallskip}
  CNF-BP with $\beta_2$& $\mathbf{\Phi}_{21}\left(10{\beta_{2}}^2{\mathbf{\Phi}_{21}}^3j-\frac{5{\beta_{2}}^2{\mathbf{\Phi}_{21}}^3\left(3{\omega_{a}}^2+\cos\left(k\right)-1\right){}j}{10{\omega_{a}}^4-8{\omega_{a}}^2-\cos\left(k\right)+5{\omega_{a}}^2\cos\left(k\right)+1}\right)$ \\
  CNF-PN with $\beta_2$& $\mathbf{\Phi}_{21}\left(10{\beta _{2}}^2{\mathbf{\Phi}_{21}}^3j-\frac{5{\beta _{2}}^2{\mathbf{\Phi}_{21}}^3\left(3{\omega_{a}}^2-2{\sin\left(k\right)}^2\right)j}{10{\omega_{a}}^4-8{\omega_{a}}^2+2{\sin\left(k\right)}^2+5{\omega_{a}}^2\left(1-2{\sin\left(k\right)}^2\right)}\right)$\\
  CNF-BP with $\gamma_1$ & $-3\gamma_{1}{\mathbf{\Phi}_{11}}^4\left({\left(1-{\mathrm{e}}^{-kj}\right)}^3+{\left(1-{\mathrm{e}}^{kj}\right)}^3\right)j$\\
  CNF-PN with $\gamma_1$ & $-3\gamma_{1}{\mathbf{\Phi}_{11}}^4{\mathrm{e}}^{-2kj}{\left(1-{\mathrm{e}}^{kj}\right)}^4j$\\
  \noalign{\smallskip}\hline
  \end{tabular}
\end{table}

\begin{figure}
    \centering
    \includegraphics[width=0.9\linewidth]{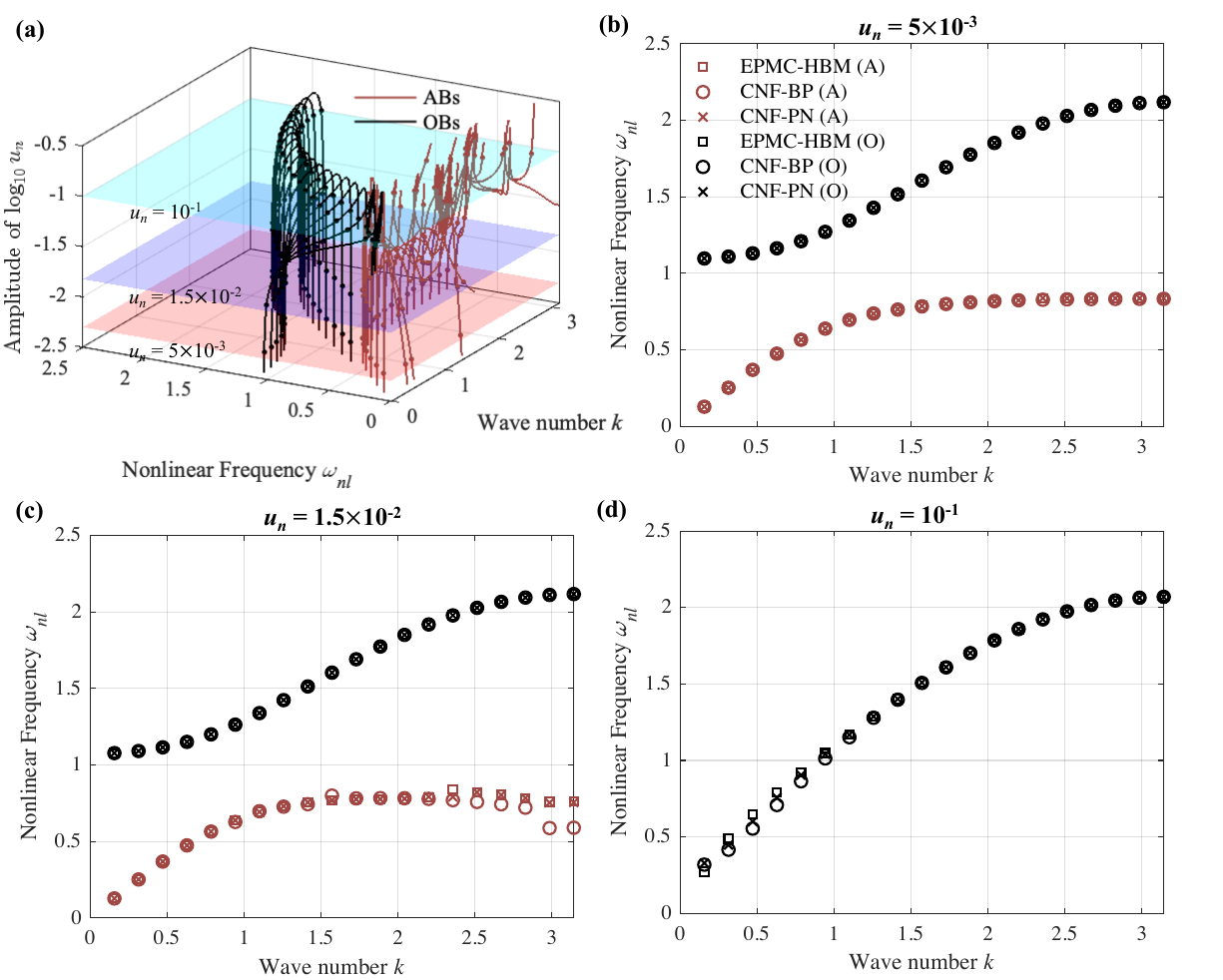}
    \caption{Nonlinear dispersion characteristics of a periodic chain with quadratic nonlinearity, $\beta_2$, in attachments: (a) Nonlinear dispersion manifold in the space of $k-\omega_{nl}-\log_{10}{u_n}$ calculated via EPMC-HBM; (b)-(d) Comparison of CNF-BP, CNF-PN, and EPMC-HBM in predicting nonlinear dispersion spectrum at different wave amplitudes, $u_n = 5\times10^{-3},1.5\times10^{-2}$, and $10^{-1}$. The red, blue, and cyan planes in (a) represent the cross-section of $u_n = 5\times10^{-3},1.5\times10^{-2}$, and $10^{-1}$. The acoustic branch in (d) is omitted. The used parameters are: $c_1 = c_2= \beta_1 =\gamma_1=\gamma_2=0$ and $\beta_2 = 2$.}
    \label{f:2}
\end{figure}

\begin{figure}
    \centering
    \includegraphics[width=0.9\linewidth]{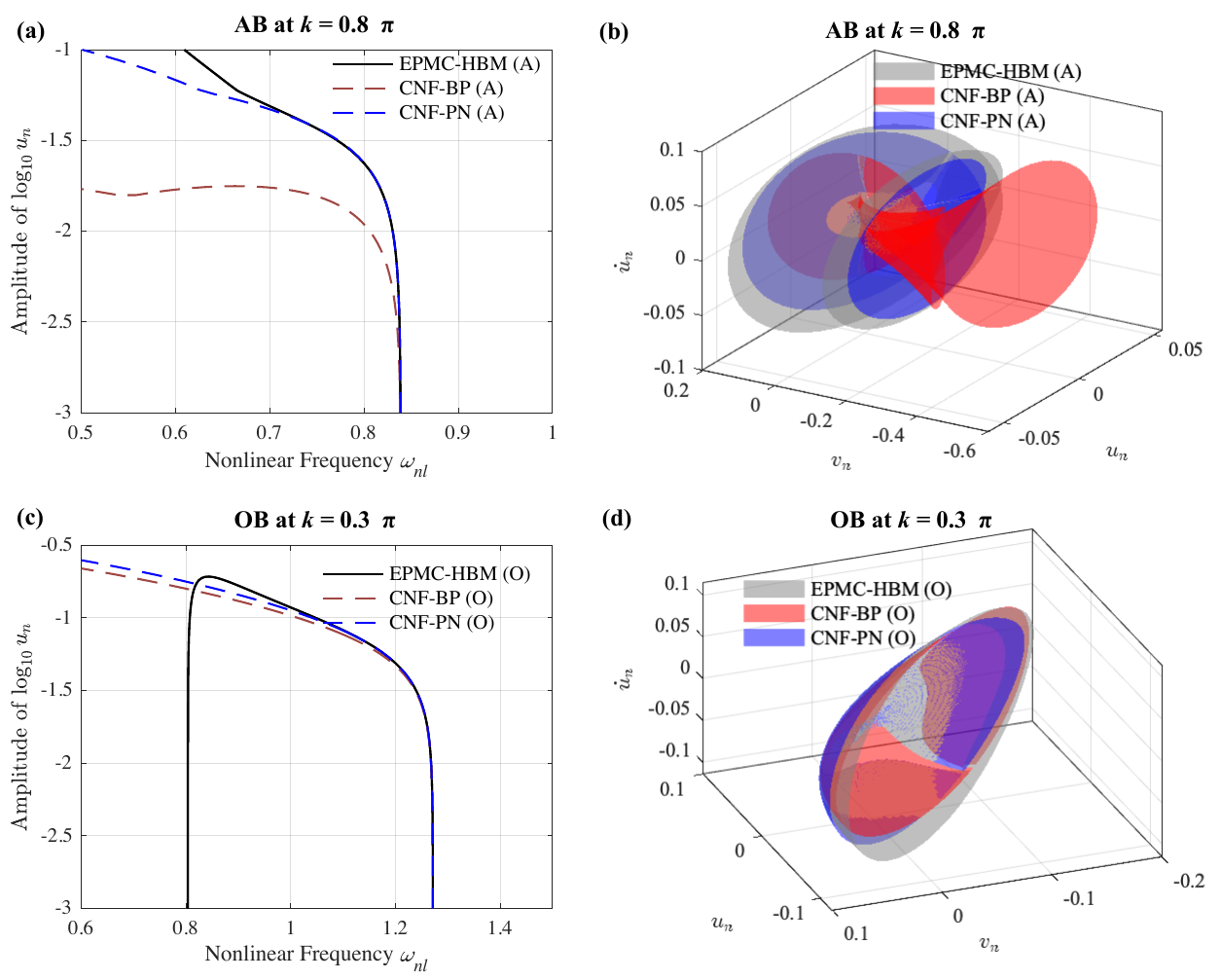}
    \caption{Backbone curves and invariant manifolds of (a)-(b) the acoustic branch (AB) at $k = 0.8\pi$ and (c)-(d) the optic branch (OB) at $k=0.3\pi$. The used parameters are: $c_1 = c_2= \beta_1 =\gamma_1=\gamma_2=0$ and $\beta_2 = 2$.}
    \label{f:3}
\end{figure}

The first example considers a nonlinear metamaterial chain with only quadratic nonlinearity in the attachments, where $c_1 = c_2= \beta_1 =\gamma_1=\gamma_2=0$ and $\beta_2 = 2$. In the calculation of both the CNF-BP and CNF-PN, the truncation order $o$ is set to 9, while the harmonic truncation $H$ in EPMC-HBM is set to 5. Based on these settings, Fig.~\ref{f:2} compares the performance of the CNF-BP and CNF-PN in predicting nonlinear dispersion behaviour. Figure~\ref{f:2}(a) provides an overview of the dispersion manifold in the $(k,\omega_{nl},\log_{10}{u_n})$ space, using EPMC-HB as reference, considering 20 wavenumbers $k = n\pi/20, n =1,2,\cdots,20$ and performing numerical continuation. Unlike an invariant manifold, which represents wave orbit variation as wave amplitude at a dispersion solution with fixed wavenumber $k$, the dispersion manifold captures the whole dispersion spectrum variations with amplitude. Results in Fig.~\ref{f:2} (a) indicate that the acoustic branch exhibits significant softening behaviour when $u_n$ exceeds $10^{-1.5}$, followed by the onset of internal resonances, which results in a very complex shape of the dispersion manifold. However, since this study excludes internal resonances between different dispersion branches in the CNF calculations, the analysis of the acoustic branch is limited to $u_n <10^{-1.5}$.

The optic branch also exhibits a softening trend with a wave amplitude limitation, which is likely due to the combined effect of quadratic nonlinearity and linear stiffness, potentially leading to two equilibria. Given that NF theory is a local method, with a validity limit in terms of amplitudes~\cite{LamarqueUP}, it cannot capture the second equilibrium, and thus, dispersion solutions near this equilibrium are omitted in the subsequent comparison of CNF-BP and CNF-PN. Based on this information, three wave amplitudes, $u_n = 5\times10^{-3},1.5\times10^{-2}$, and $10^{-1}$, are selected in Figures~\ref{f:2}(b)-(d) to compare the accuracy of CNF-BP and CNF-PN in predicting the dispersion spectrum at different wave amplitudes. The results show that both methods yield similar predictions for nearly linear cases ($u_n = 5\times10^{-3}$). However, as the wave amplitude increases, frequency shifts occur: the acoustic branch shifts at larger wavenumbers, while the optic branch shifts at smaller wavenumbers. In both cases, the CNF-PN demonstrates better agreement with the continuation results. This suggests that the PN strategy generally offers improved accuracy. However, it is worth mentioning that the parametrisation method with PN strategy remains a local technique and becomes ineffective when the wave amplitude continues to increase and exceeds its validity limit. 

To give a more quantitative comparison, Fig.~\ref{f:3} presents the backbone curves (natural frequency variation with wave amplitude) and invariant manifolds of the acoustic branch at a large wavenumber ($k=0.8\pi$) and the optic branch at a small wavenumber ($k = 0.3\pi$). For the acoustic branch, the CNF-PN’s prediction closely matches the HBM results up to $u_n = 10^{-1.3}$, whereas the CNF-BP exhibits significant deviations even at relatively small wave amplitudes, such as $10^{-2}$. The discrepancy arises from the stark difference in the nonlinear coefficients of the CNF. Indeed, a close inspection of the computed coefficients reveals that the first nonlinear correction is computed as $\varpi_1 = -11.37j$ for CNF-PN against $\varpi_1 = -34.92j$ for CNF-BP, using the formula in Table~\ref{tab:2}. Additionally, the invariant manifold predicted by CNF-BP  and shown in Fig.~\ref{f:3}(b) exhibits significant deviations, whereas CNF-PN aligns closely with continuation results, even capturing manifold folding—a hallmark of non-fundamental harmonic generation. These folding dynamics imply the significant non-trivial resonant terms of $\boldsymbol{\alpha}(p,q) \notin [n+1,n]$ or $[n,n+1]$, which are mishandled by CNF-BP due to its inability to account for wave-wave interaction in Eq.~\eqref{eq:6}. For the optic branch, where energy predominantly concentrates in the fundamental harmonic (no manifold folding observed in Fig.~\ref{f:3}(d)), both CNF-BP ($\varpi_1=-22.31j$) and CNF-PN ($\varpi_1=-21.79j$) yield comparable backbone curves and invariant manifolds. Nevertheless, CNF-PN achieves marginally higher accuracy and a larger validity limit, particularly in resolving amplitude-dependent frequency shifts. However, as shown in Figures~\ref{f:3}(c) and (d), the CNF-PN still outperforms the CNF-BP, further demonstrating its superior accuracy in capturing nonlinear dispersion characteristics under various conditions.

\begin{figure}
    \centering
    \includegraphics[width=0.9\linewidth]{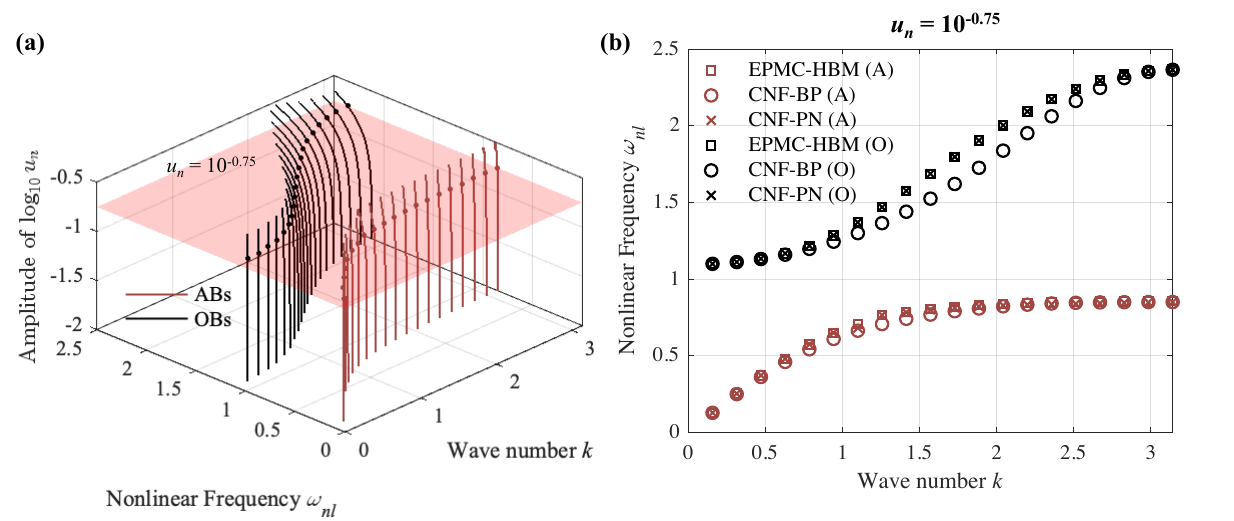}
    \caption{Nonlinear dispersion characteristics of a periodic chain with cubic nonlinearity, $\gamma_1$, between host lattices: (a) Nonlinear dispersion manifold in the space of $k-\omega_{nl}-\log_{10}{u_n}$ calculated via EPMC-HBM; (b) Comparison of CNF-BP, CNF-PN, and EPMC-HBM in predicting nonlinear dispersion spectrum at wave amplitude, $\max{(u_n)} = 10^{-0.75}$. The red plane in (a) represents the cross-section of $u_n = 5\times10^{-0.75}$. The used parameters are: $c_1 = c_2= \beta_1 =\beta_2=\gamma_2=0$ and $\gamma_1 = 3$.}
    \label{f:4}
\end{figure}

\begin{figure}
    \centering
    \includegraphics[width=0.9\linewidth]{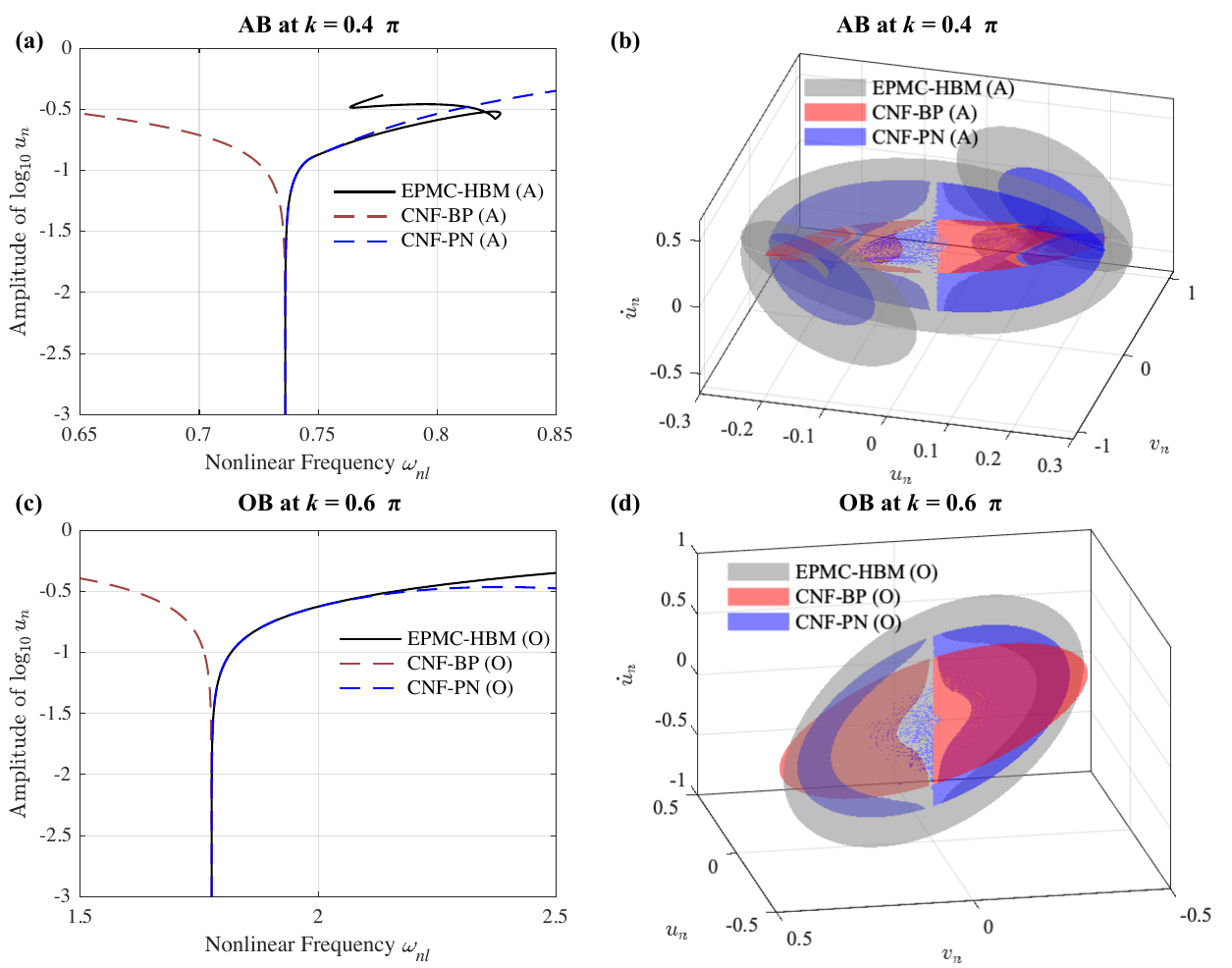}
    \caption{Backbone curves and invariant manifolds of (a)-(b) the acoustic branch (AB) at $k = 0.4\pi$ and (c)-(d) optic branch (OB) at $k=0.6\pi$. The used parameters are: $c_1 = c_2= \beta_1 =\beta_2=\gamma_2=0$ and $\gamma_1 = 3$.}
    \label{f:5}
\end{figure}

Fig.~\ref{f:4} investigates the limitations of the BP strategy in addressing nonlinearities in host oscillators by analysing the dispersion characteristics of a conservative periodic chain with cubic nonlinearity, $\gamma_1$, between host lattices. In this case, $\gamma_1 = 3$, while all other parameters are set to zero ($c_1=c_2=\beta_1=\beta_2=\gamma_2 = 0$). Figure~\ref{f:4}(a) presents the 3D dispersion manifold obtained using EPMC-HBM, which reveals a hardening behaviour in both the acoustic and optic branches. Figure~\ref{f:4}(b) compares the CNF-BP and CNF-PN predictions for wave amplitudes $u_n = 10^{-0.75}$, demonstrating that the PN strategy yields dispersion spectra that closely align with EPMC-HBM results. In contrast, the CNF-BP exhibits significant discrepancies, particularly within the ranges $0.2\pi<k<0.7\pi$ for the acoustic branch and $0.2\pi<k<0.9\pi$ for the optic branch. These discrepancies correlate with the nonlinear coefficient $\varpi_1$ in Table~\ref{tab:2}, where CNF-BP and CNF-PN yield identical $\varpi_1$ only at $k = 0$ or $\pi$, but diverge across intermediate wavenumbers. To further examine these discrepancies, Figure~\ref{f:5} provides a detailed analysis of the dispersion solutions and invariant manifolds at fixed wavenumbers, specifically for the acoustic branch at $k=0.4\pi$ and the optic branch at $k=0.6\pi$. The results indicate that the CNF-BP exhibits substantial deviations from EPMC-HBM predictions in both backbone curves and invariant manifolds. Specifically, the BP strategy provides $\varpi_1 = -0.38j$ for the acoustic branch and $\varpi_1 = -1.67j$ for the optic branch, leading to inaccuracies in predicting the hardening/softening behaviour. In contrast, CNF-PN achieves significantly higher accuracy, with $\varpi_1 = 0.24j$ for the acoustic branch and $\varpi_1 = 4.37j$ for  the optic branch, closely aligning with EPMC-HBM results until $u_n \simeq 10^{-0.8}$ for the acoustic branch or $u_n \simeq 10^{-0.5}$ for the optic branch.

\subsection{Effect of Truncation Order}
\label{sec:4.2}
\begin{figure}
    \centering
    \includegraphics[width=0.9\linewidth]{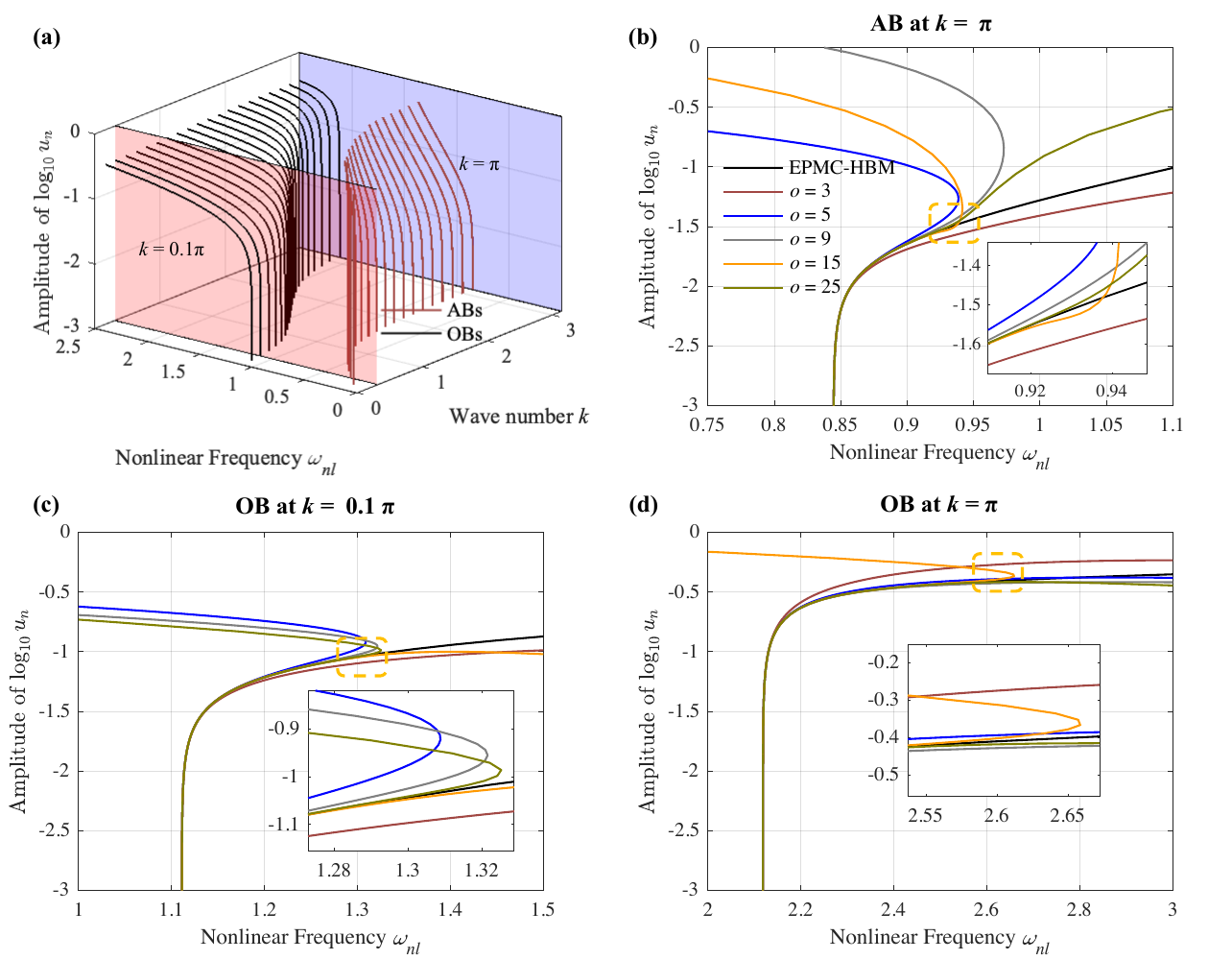}
    \caption{Nonlinear dispersion characteristics of a periodic chain with cubic nonlinearity, $\gamma_2$, in the attachments: (a) Nonlinear dispersion manifold in the space of $(k,\omega_{nl},\log_{10}{u_n})$ numerically obtained via EPMC-HBM; (b)-(d) Comparison of CNF-PN with truncation orders $o = 3,5,9,15,$ and $25$ in predicting the backbone curves of (b) the acoustic branch (AB) at $k = \pi$, and the optic branch (OB) at (c) $k = 0.1\pi$ and (d) $k=\pi$. The red and blue planes in (a) represent the cross-section of $k = 0.1\pi$ and $\pi$ selected to plot the backbone curves. The used parameters are: $c_1 = c_2= \beta_1 =\beta_2=\gamma_1=0$ and $\gamma_2 = 3$.}
    \label{f:6}
\end{figure}

Section~\ref{sec:4.1} demonstrated the advantages of the PN strategy; however, the effect of the truncation order~$o$ and the convergence of the CNF expansions warrants further investigation. Figure~\ref{f:6}(a) presents the dispersion manifold of an undamped metamaterial chain with cubic nonlinearity only ($\gamma_2=3$, with all other parameters set to zero) obtained using EPMC-HBM as a reference numerical solution. The results indicate a hardening behaviour in both the acoustic and optic branches. To assess the impact of truncation order on accuracy and validity limits, three dispersion solutions exhibiting significant bending are examined: the acoustic branch at a large wavenumber ($k = \pi$) and the optic branches at both a small ($k=0.1\pi$) and a large wavenumber ($k=\pi$). CNF-PN predictions for the backbone curves are compared across different truncation orders ($o = 3, 5, 9, 15, 25$). 

The results clearly indicate that, by increasing the order of the asymptotic expansion in the CNF-PN development, a better accuracy is obtained, but in a limited amplitude range which corresponds to the validity limit of the method. This awaited result is the consequence of the local nature of the normal form approach~\cite{RN37}, which converges to the exact solution but only when the amplitude does not exceed the validity range~\cite{RN41,LamarqueUP}. In the present case, the validity limit of the CNF expansion can be estimated from the numerical results at $u_n = 10^{-1.5}$ for the acoustic branch with $k=\pi$ shown in Fig.~\ref{f:6}(b), $u_n = 10^{-1}$ and $u_n = 10^{-0.4}$  for the optic branch respectively computed at $k=0.1\pi$ and $k=\pi$ shown in Figs.~\ref{f:6}(c-d). Note that each of these cases corresponds to a frequency shift along the backbone curve of about 20\%. Once the validity limit is reached, a sharp departure from the reference solution is generally observed, related to the use of high-order polynomials. Nevertheless, within the validity bound, a perfect convergence is at hand when increasing the orders, such that those asymptotic developments are sometimes referred to as {\em exact}, referring to their ability to converge to the correct solution within the valid amplitude range.


\subsection{Comparison with other methods}
\label{sec:4.3}

\begin{figure}
    \centering
    \includegraphics[width=0.9\linewidth]{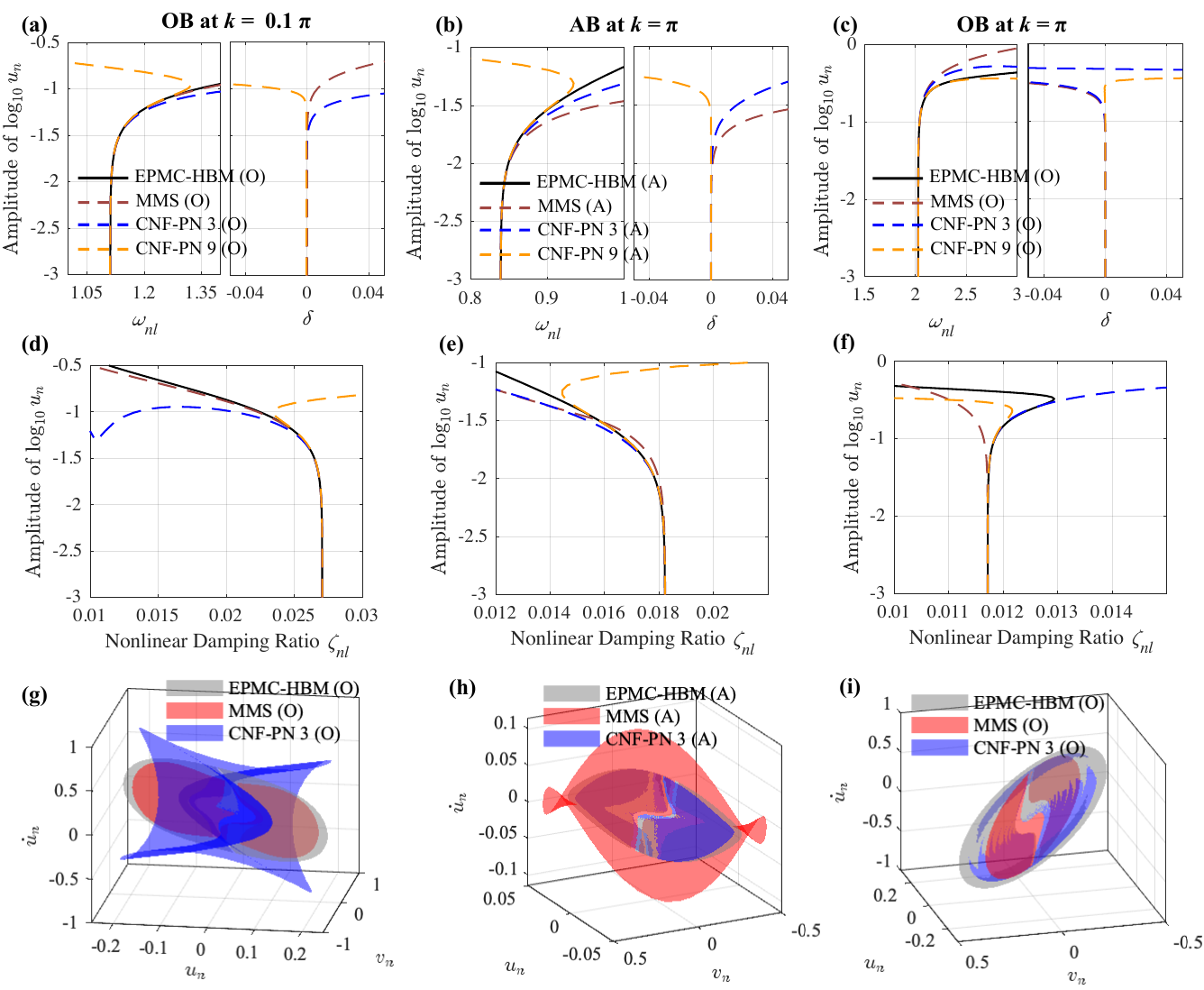}
    \caption{Nonlinear dispersion characteristics of an underdamped periodic chain with cubic nonlinearity, $\gamma_2$, in attachments: right (a)-(c) Optic backbone (OB) at $k = 0.1\pi$, acoustic (AB) and optic backbones (OB) at $k = \pi$; left (a)-(c) demonstrate the relative difference $\delta$ between the backbonecurves predicted by MMS, CNF-PN 3, CNF-PN 9 and that of EPMC-HBM; (d) Optic damping ratio at $k = 0.1\pi$; (e)-(f) Acoustic and optic damping ratios at $k=\pi$; (g) Optic invariant manifold at $k = 0.1\pi$; (h)-(i) Acoustic and optic invariant manifolds at $k = \pi$. The used parameters are: $c_1 = 5\times10^{-3},c_2=2\times10^{-2},\beta_1 =\beta_2=\gamma_1=0$ and $\gamma_2 = 3$. The relative difference $\delta$ is given by $(\omega_{nl}-\omega_{nl,HBM})/\omega_{nl,HBM}$, where the $\omega_{nl,HBM}$ is from the EPMC-HBM, as for $\omega_{nl}$ is obtained via MMS, CNF-PN 3, or CNF-PN 9.}
    \label{f:8}
\end{figure}

\begin{figure}
    \centering
    \includegraphics[width=0.45\linewidth]{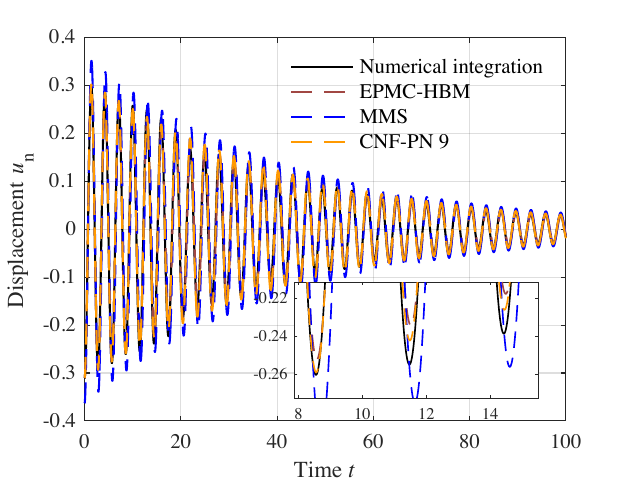}
    \caption{Comparison of HBM, MMS, and CNF-PN ($o=9$) in predicting wave attenuation in an underdamped periodic chain with cubic nonlinearity in attachments. The numerical integration results are obtained by collecting the response of the 100-th lattice in the annular chain composed of 100 lattices. The initial wave locates the optic branch of EPMC-HBM with amplitude and wavenumber $u_n = 10^{-0.5}$ and $k = 0.8\pi$. The EPMC-HBM uses the expression, $\mathbf{x}_{n+i} = a_0e^{-\zeta_{nl}\omega_{nl}t}\left[\mathbf{A}_0+\sum_{m=1}^{H}\mathbf{A}_me^{j(\omega_{nl}t-ki)}\right] + c.c.$~\cite{RN11} to predict the wave attenuation process. The amplitude dependence of some variables is obtained by interpolating the continuation result of EPMC-HBM. As for MMS and CNF-PN, their wave attenuation is calculated via Eq.~\eqref{eq:24} and~\eqref{eq:43}. The used parameters are: $c_1 = 5\times10^{-3},c_2=2\times10^{-2},\beta_1 =\beta_2=\gamma_1=0$ and $\gamma_2 = 3$.}
    \label{f:9}
\end{figure}

\begin{figure}
    \centering
    \includegraphics[width=0.9\linewidth]{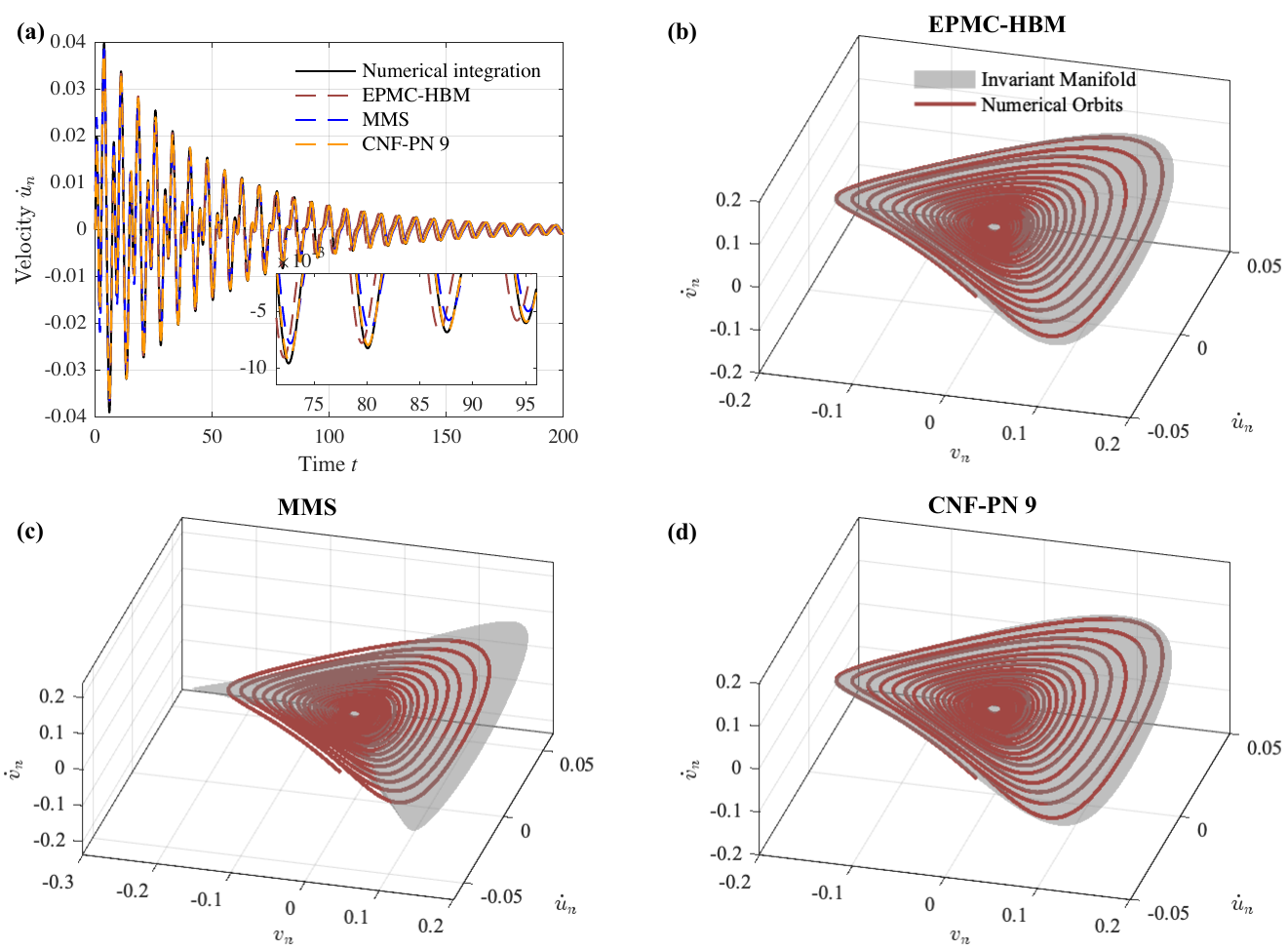}
    \caption{(a) Comparison of HBM, MMS, and CNF-PN ($o=9$) in predicting wave attenuation in an underdamped periodic chain with various nonlinearities. The numerical integration results are obtained by collecting the response of the 100-th lattice in the annular chain composed of 100 lattices. The initial wave locates the optic branch of EPMC-HBM with amplitude and wavenumber $u_n = 10^{-1.4}$ and $k = 0.8\pi$. The EPMC-HBM uses the expression, $\mathbf{x}_{n+i} = a_0e^{-\zeta_{nl}\omega_{nl}t}\left[\mathbf{A}_0+\sum_{m=1}^{H}\mathbf{A}_me^{j(\omega_{nl}t-ki)}\right] + c.c.$ to predict the wave attenuation process. The amplitude dependence of some variables is obtained by interpolating the continuation result of EPMC-HBM. As for MMS and CNF-PN, their wave attenuation is calculated via Eq.~\eqref{eq:24} and~\eqref{eq:43}. (b)-(d) Wave attenuation orbits on the invariant manifolds predicted by HBM, MMS, and CNF-PN. The used parameters are: $c_1 = 5\times10^{-3},c_2=2\times10^{-2},\beta_1 = 1, \beta_2=-1, \gamma_1=2$ and $\gamma_2 = 3$. Note: the $u_n$ are avoided because the quadratic nonlinearity and initial condition may bring rigid-body motion in the numerical integration of the annular chain.}
    \label{f:10}
\end{figure}

For the undamped case, the previous two sections have provided a detailed discussion, demonstrating both the advantages of the CNF-PN strategy and the effect of truncation order. However, its performance for underdamped periodic chains, as well as its advantages and limitations compared to classical methods such as the harmonic balance method (EPMC-HBM) and the analytical method of multiple scales (MMS), still warrants further investigations. Therefore, this section introduces damping terms, setting $c_1 = 5\times10^{-3}$ and $c_2 = 2\times10^{-2}$, into two types of periodic chains. Example 1 considers an underdamped chain with cubic nonlinearity in attachments, $\gamma_2=3$; while example 2 tackles the case of an underdamped chain with various nonlinearities, $\beta_1=1,\beta_2=-1,\gamma_1=2$, and $\gamma_2=3$. The comparisons between CNF-PN, HBM-EPMC and MMS, are drawn out on their ability to correctly predict: the backbone curves of dispersion solutions (Example 1), the nonlinear damping ratio (Example 1), the invariant manifolds (Examples 1 and 2), as well as the wave attenuation (Examples 1 and 2). Notably, an annular chain composed of 100 nonlinear lattices is included in each example to provide numerical integration results. 

\begin{table}[h!]
  \centering
  \begin{tabular}{ll}
  \hline\noalign{\smallskip}
  Coefficient & Expression \\
  \noalign{\smallskip}\hline\noalign{\smallskip}
  $\varpi_1$ & $-3\gamma_2\mathbf{\Phi}_{21}^3\bar{\mathbf{\Phi}}_{21}$ \\
  $\varepsilon^2D_2\psi$ & $\frac{3\rho_{m}^2\gamma_2\mathbf{\Phi}_{m2,0}^4}{8\omega_{0}}$\\
  \noalign{\smallskip}\hline
  \end{tabular}
  \caption{The expressions, $\varpi_1$, of CNF-PN and $\varepsilon^2D_2\psi$ for nonlinear metamaterial chain with $\gamma_2$. Note: the acoustic and optic CNF-PN share the same $\varpi_1$ in this example.}
    \label{tab:3}       
\end{table}


To start with, an analytical comparison in terms of the coefficients of the backbone curve predicted by both the MMS and the CNF-PN at the same order three, is presented for Example 1. The coefficients of the backbone curve computed using the CNF-PN are given in Eqs.~\eqref{eq:24}, while those provided by deploying the method of multiple scales are recalled in~\ref{sec:A2}. Focusing on the first nonlinear correction, {\emph i.e.} the term at order $o=3$, Table~\ref{tab:3} details the expression of the associated coefficient, namely $\varpi_1$ for CNF-PN and $\varepsilon^2D_2\psi$ for MMS, for the specific case of Example 1. In the reported expressions, $\mathbf{\Phi}_{m,0}$ ($\mathbf{\Phi}_{m2,0}$ denotes its second element) and $\rho_m$ represent, respectively, the conservative real-valued wave shape and the amplitude associated with the acoustic or optic dispersion branches, as defined for the MMS in~\ref{sec:A2}. In contrast, $\mathbf{\Phi}_{1}$ ($\mathbf{\Phi}_{21}$ is its second element) denotes the complex-valued wave shape of the selected dispersion branch, which is applicable to both damped and undamped cases. In the conservative case, the two wave shapes and the amplitude $\rho_m$ satisfy the following relationships (detailed explanation in~\ref{sec:A2}):
$\mathbf{\Phi}_{1} = \frac{\mathbf{\Phi}_{m,0}}{\sqrt{2\omega_{0}j}},\rho = \sqrt{2\omega_{0}}\rho_m$. In this particular case, MMS and CNF-PN with $o=3$ yield identical backbone formula, given by: $\omega_{nl} = \omega_{0} + \frac{3\gamma_2\mathbf{\Phi}_{m2,0}^4}{8\omega_{0}}\rho_m^2$. However, when returning back to physical coordinates, a difference is still at hand: since  $u_n$ represents the wave amplitude, the differences brought about by the nonlinear mappings lead to discrepancies in backbone predictions, as illustrated in the following Fig.~\ref{f:8}, and as already commented for different cases of nonlinear oscillations in~\cite{RN46}. For the underdamped case, the MMS still yields the backbone curve's expression and nonlinear mapping in the same form as the undamped system, and therefore fails to capture the coupling between damping and nonlinearity. In contrast, the CNF-PN overcomes this limitation and provides an exact prediction of damping effects.

A numerical comparison of different computed backbone curves is shown in Fig.~\ref{f:8} for example 1. Three different subcases have been selected for their representativeness of the available results: (i) the optic branch at a small wavenumber $k=0.1\pi$; (ii) the acoustic branch at a large wavenumber $k=\pi$; and (iii) the optic branch at a large wavenumber $k=\pi$. The general trend must be understood in light of the local nature of the asymptotic expansion provided by the CNF-PN solution, which has a validity limit in terms of amplitude~\cite{LamarqueUP,RN41,groletNFnonpoly,RN46}: as long as the validity limit is not reached, then the asymptotic expansion converges to the correct solution, providing a very accurate approximation of the exact backbone curve when increasing the order. However, once the validity limit is reached, the departure from the exact solution is all the more severe as a high-order series has been computed. On the other hand, the MMS solution, limited to order 3, does not provide convergence and can be close or far from the exact solution, with no possibility of increasing the order since it involves difficult calculations.


For most of the computed results, the CNF-PN with order 3 gives better results than the MMS solution, as illustrated in Figs.~\ref{f:8}(b,c) This is due to the most accurate computation of the physical displacement thanks to the nonlinear mapping, as commented before. However, in a few cases, the opposite happens, as illustrated in Fig.~\ref{f:8}(a). Nevertheless, one can observe that increasing the order gives a better accuracy inside the range of convergence of the CNF-PN. In the case reported in Fig.~\ref{f:8}(a), the validity limit can be approximated to $u_n\approx 10^{-1.1}$. In order to better quantify this accuracy, the right panels in Figs.~\ref{f:8}(a,b,c) display the difference $\delta$ between the backbone curve computed numerically with EPMC-HBM, and taken as reference, and either CNF-PN or MMS. One can conclude that the valdity limit is case-dependent, but as long as the limit is not reached, the CNF-PN solutions converge to the exact. On the other hand, MMS solutions are in general not accurate enough, with no possibility for the analyst to increase the quality of the result. The fact that MMS give better approximations in a few cases is thus interpreted as incidental, whereas CNF-PN offers a solution with guaranteed convergence and a clearly established limit defining the region of validity. 


The second and third rows of Fig.~\ref{f:8} confirmed the previous analyses by highlighting the different approximations in terms of nonlinear damping ratio (second row) and invariant manifolds (third row). In particular, the invariant manifold can have a complex shape, representing an important nonlinearity in the relationship between the normal and physical coordinates. In this case, it is thus expected that the CNF-PN strategy rapidly outperforms the MMS solution, due to its treatment of the amplitudes. On the other hand, for cases where the manifold is flat and the relationship between normal and physical coordinates is close to linear, this disparity is decreasing.

An interesting observation is finally reported in Fig.~\ref{f:8}(c,f), for the case of the optic branch at $k=\pi$: 
the damping ratio predicted by the CNF-PN with $o=9$ deviates more from EPMC-HBM than that predicted by the CNF-PN with $o=3$ when $u_n$ approaches $10^{-0.5}$, despite the former's superior backbone prediction within this range. This phenomenon suggests that either the CNF-PN solution or the EPMC-HBM computation may produce inaccurate predictions for the damping ratio. To investigate this discrepancy, Figure~\ref{f:9} presents numerical integration results for an annular chain with 100 lattices as a reference. The wave attenuation predicted by EPMC-HBM, MMS,  CNF-PN with $o=9$, and direct numerical time integration, are compared. The results indicate that the CNF-PN aligns more closely with the numerical results than EPMC-HBM, while MMS performs poorly. This implies that, within its effective range, the CNF-PN is capable of providing the most accurate predictions for various dispersion characteristics, including the damping ratio, in underdamped periodic chains. 
The inaccurate prediction of the damping characteristics provided by the EPMC-HBM should be attributed to the strong assumptions that are made by the technique itself. Indeed, the method introduces an artificial negative damping force that is proportional to mass, to balance the internal damping. But such is simple strategy can properly work out for simple damping cases and cannot handle all the complexities provided by the losses. Hence the results found here illustrate this limitation.

Figure~\ref{f:10} further validates these findings and assesses the accuracy of the invariant manifolds predicted by EPMC-HBM, MMS, and the CNF-PN using a more general nonlinear periodic chain ($\beta_1=1,\beta_2=-1,\gamma_1=2$, and $\gamma_2=3$). The results demonstrate that the CNF-PN continues to provide the best predictions, as shown in Fig.~\ref{f:10} (a), followed by EPMC-HBM and then MMS, even in the presence of a generated second harmonic. Figures~\ref{f:10} (b)-(d) illustrate the wave decay process on the invariant manifolds predicted by EPMC-HBM, MMS, and the CNF-PN with $o=9$. The numerical orbits align closely with the invariant manifolds predicted by EPMC-HBM and the CNF-PN, while showing slight deviations from those predicted by MMS. This further corroborates the accuracy of the invariant manifolds derived from the three methods.

\section{Conclusion}
\label{sec:5}

This study is devoted to using the normal form embedded in the direct parametrisation method for invariant manifolds, in order to derive the nonlinear dispersion curves of an infinite periodic chain with internal resonators and various nonlinearities. Two methods have been proposed, characterised by the order in which the two main assumptions ({\it i.e.} the nonlinear mapping and Bloch's periodic assumption) are inserted in the calculation scheme. In the CNF-BP method, Bloch's assumption is first applied to the physical coordinates. On the other hand, in the CNF-PN strategy, the nonlinear mapping and normal form of the model is first derived, while the periodic wave assumption is then applied to the normal coordinates.


It has been shown that the BP strategy exhibits limitations in handling non-fundamental harmonics and nonlinearities in host masses, leading to significant deviations in predictions for some cases, such as a system with quadratic nonlinearities or nonlinear coupling terms between lattices. In contrast, the PN strategy overcomes these limitations by incorporating periodic constraints on the normal coordinates, enabling a more robust and accurate representation of nonlinear wave propagation in various nonlinear metamaterials.



The comparison with EPMC-HBM and the method of multiple scales highlighted the CNF-PN’s capability to accurately predict wave attenuation and damping ratios in underdamped systems. While EPMC-HBM provides highly accurate results for undamped cases, its performance degrades in the presence of damping, where the CNF-PN excels within the effective range. As a general rule, CNF-PN predictions have been found to be accurate and convergent in any of the cases treated, as long as the validity limit of the asymptotic expansion is not reached. For practical applications, a truncation order of $o=9$ strikes a balance between computational efficiency and accuracy, making it suitable for designing and optimising nonlinear metamaterials.

In this contribution, internal resonances between acoustic and optic branches have not been taken into account. However, complex resonance scenarios have been shown to occur for such resonant metamaterial lattice, as shown for instance in~\cite{RN11}. A natural extension to the present work is thus to develop the normal form for these resonant cases, which could be done in case-by-case studies with the different normal forms describing each of the possible resonances.

In conclusion, the CNF-PN strategy represents a significant advancement in the analysis of infinite periodic and underdamped nonlinear chains, offering a powerful local technique for understanding and designing nonlinear metamaterials. This method has been incorporated into a numerical open-access package, \emph{MORFE\_metamaterial}, enabling numerical computation of higher-order CNF-PN for a given wave number, as well as the rapid retrieval of the third-order truncated symbolic CNF-PN.

\section*{Acknowledgements}
This work is supported by the State Key Laboratory of Micro-Spacecraft Rapid Design and Intelligent Cluster (MS01240117), National Natural Science Foundation of China (Nos. 12132010, 12021002), Natural Science Foundation of Tianjin (No. 23JCQNJC01800), and China Scholarship Council (202406250148).

\section*{Data availability}

The source code for running the simulations is provided as a matlab package that can be downloadable from the github page of the \texttt{MORFE} project: \textit{https://github.com/MORFEproject}. More specifically, the open source codes are grouped in the repository \emph{MORFE\_metamaterial} which can be accessed at \textit{https://github.com/MORFEproject/MORFE\_metamaterial}.

\section*{Declaration of Interest Statement}
The authors declare that they have no known competing financial interests or personal relationships that could have appeared to inﬂuence the work reported in this paper.

\appendix
\section{Derivation of the expressions for $\dot{\mathbf{x}}_n, \dot{\mathbf{x}}_{n\pm1}$, and $\dot{\mathbf{y}}_n$}
\label{sec:Add}
This section presents the detailed derivation of the expressions for $\dot{\mathbf{x}}_n, \dot{\mathbf{x}}_{n\pm1}$, and $\dot{\mathbf{y}}_n$ as given in Eq.~\eqref{eq:12}. Starting with $\dot{\mathbf{x}}_n$, it can be expressed as
\begin{equation}
    \label{eq:an1}
    \dot{\mathbf{x}}_n = \frac{\partial\mathbf{x}_n}{\partial \mathbf{z}_n}\dot{\mathbf{z}}_n=\nabla_z\hat{\mathbf{\Psi}}(\mathbf{z}_n)\hat{\mathbf{f}}(\mathbf{z}_n),
\end{equation}
where $\hat{\mathbf{f}}(\mathbf{z}_n)$ is defined in Eq.~\eqref{eq:11}, and the gradient $\nabla_z\hat{\mathbf{\Psi}}(\mathbf{z}_n)$ takes the form $\nabla_z\hat{\mathbf{\Psi}}(\mathbf{z}_n)=[\frac{\partial \hat{\mathbf{\Psi}}(\mathbf{z}_n)}{\partial z_{1}},\frac{\partial \hat{\mathbf{\Psi}}(\mathbf{z}_n)}{\partial z_{2}}]$. Making use of the expression of $\hat{\mathbf{\Psi}}(\mathbf{z}_n)$, each partial derivative $\frac{\partial \hat{\mathbf{\Psi}}(\mathbf{z}_n)}{\partial z_{ns}}$ is computed as
\begin{equation}
    \label{eq:an2}
    \frac{\partial \hat{\mathbf{\Psi}}(\mathbf{z}_n)}{\partial z_{s}} = \mathbf{\Phi}_s + \sum_{p=2}^{o}\sum_{q=1}^{q_p}\alpha_s(p,q)\hat{\mathbf{\Psi}}^{(p,q)}\mathbf{z}_n^{\boldsymbol{\alpha}(p,q)-\boldsymbol{e}_s},
\end{equation}
where $\mathbf{e}_s = [1;0]$ for $s=1$, or $[0;1]$ for $s=2$. Taking $s = 1$ as example, the $\mathbf{z}_n^{\boldsymbol{\alpha}(p,q)-\boldsymbol{e}_s} = z_{1}^{\alpha_1-1}z_{2}^{\alpha_2}$. Substituting Eq.~\eqref{eq:an2} and the expression of $\hat{\mathbf{f}}(\mathbf{z}_n)$ in Eq.~\eqref{eq:11} into Eq.~\eqref{eq:an1} yields the same expression of $\dot{\mathbf{x}}_n$ as that in Eq.~\eqref{eq:12}.

In a similar manner, the $\dot{\mathbf{x}}_{n\pm1}$ is given by
\begin{equation}
\label{eq:an3}
   \dot{\mathbf{x}}_{n\pm1} = \frac{\partial\mathbf{x}_{n\pm1}}{\partial \mathbf{z}_n}\dot{\mathbf{z}}_n=\nabla_z\hat{\mathbf{\Psi}}(\mathbf{E}_{\pm1}\mathbf{z}_n)\hat{\mathbf{f}}(\mathbf{z}_n), 
\end{equation}
where $\nabla_z\hat{\mathbf{\Psi}}(\mathbf{E}_{\pm1}\mathbf{z}_n) = [\frac{\partial \hat{\mathbf{\Psi}}(\mathbf{E}_{\pm1}\mathbf{z}_n)}{\partial z_{1}},\frac{\partial \hat{\mathbf{\Psi}}(\mathbf{E}_{\pm1}\mathbf{z}_n)}{\partial z_{2}}]$ and each partial derivative is given by 
\begin{equation}
    \label{eq:an4}
     \frac{\partial \hat{\mathbf{\Psi}}(\mathbf{E}_{\pm1}\mathbf{z}_n)}{\partial z_{ns}} = \mathbf{\Phi}_s\mathbf{E}_{\pm1s} + \sum_{p=2}^{o}\sum_{q=1}^{q_p}\alpha_s(p,q)e^{\pm jk(\alpha_2(p,q)-\alpha_1(p,q))}\mathbf{z}_n^{\boldsymbol{\alpha}(p,q)-\boldsymbol{e}_s}
\end{equation}
with $\mathbf{E}_{\pm1s} = e^{\mp jk}$ for $s=1$, and $e^{\pm jk}$ for $s=2$. Substitution of Eq.~\eqref{eq:an4} and $\hat{\mathbf{f}}(\mathbf{z}_n)$ into Eq.~\eqref{eq:an3} results in the expression of $\dot{\mathbf{x}}_{n\pm1}$ consistent with Eq.~\eqref{eq:12}.

Finally, the derivation of $\dot{\mathbf{y}}_n$ follows the same procedure as that of $\dot{\mathbf{x}}_n$, with the only modification being the replacement of $\hat{\mathbf{\Psi}}$ by $\hat{\mathbf{\Upsilon}}$.
\section{EPMC-HBM}
\label{sec:A1}
For the EPMC-HBM the displacement in Eq.~\eqref{eq:1} is assumed as a Bloch wave expansion consisting of integer multiples of the frequency and wavenumber, expressed as~\cite{RN11,RN33}
\begin{equation}
    \label{eq:25}
    \mathbf{x}_{n+i} = a\left[\mathbf{A}_0 + \sum_{m=1}^{H}{\mathbf{A}_m e^{mj(\omega_{nl} t-ki)}}\right]+c.c.
\end{equation}
Substituting Eq.\eqref{eq:25} into Eq.\eqref{eq:1}, adding an artificial damping term~\cite{Krack_EPMC}, $-2\zeta_{nl}\omega_{nl} \mathbf{M}\dot{\mathbf{x}}_n$, to counterbalance the dissipated energy by damping and performing harmonic balance in the time domain, the residual equation is derived as
\begin{equation} 
\label{eq:26}
\begin{aligned}
    \mathbf{R}_{c}=a\left[-\omega_{nl}^{2} \nabla^{2} \otimes \mathbf{M}+j\omega_{nl}\left(\nabla\otimes\left(\mathbf{C}_0-2\zeta_{nl}\omega_{nl}\mathbf{M}\right)\right.\right.\\
    \left.\left.+2\nabla\nabla_k\otimes\mathbf{C}_{-1}\right)+\mathbf{I} \otimes \mathbf{K}_{0}+2 \nabla_{k} \otimes \mathbf{K}_{-1}\right] \hat{\mathbf{A}}
    +\mathbf{F}_{n H}=0,
\end{aligned}
\end{equation}
where
\begin{subequations}
\begin{align}
    \nabla &= \mathrm{diag} \left(\left[\begin{array}{lllll}
    0 & 1 & 2 & \cdots & H
    \end{array}\right]\right), \\
    \nabla_{k} &= \mathrm{diag} \left(\left[\begin{array}{llll}
    1 & \cos k & \cdots & \cos H k
    \end{array}\right]\right), \\
    \hat{\mathbf{A}}&=\left[\mathbf{A}_{0}^{\mathrm{T}}, \mathbf{A}_{1}^{\mathrm{T}}, \cdots, \mathbf{A}_{H}^{\mathrm{T}}\right]^{\mathrm{T}}, \\
    \mathbf{F}_{n H}&=\left[\mathbf{f}_{n 0}^{\mathrm{T}}, \mathbf{f}_{n 1}^{\mathrm{T}}, \cdots, \mathbf{f}_{n H}^{\mathrm{T}}\right]^{\mathrm{T}}.
\end{align}
\end{subequations}
The notation $\otimes$ denotes the Kronecker Product~\cite{RN11,Wang_dNNM}. The nonlinear forces in Eq.~\eqref{eq:1}, $\mathbf{F}_n,\mathbf{F}_{nb}$, are handled using the fast Fourier transform to obtain $\mathbf{f}_{nH}$~\cite{Krack_book}. Next, two constraint equations are introduced,
\begin{equation}
  \label{eq:28}
  \mathbf{R}_{1}=\mathrm{Re}\left(\sum_{m=0}^{H} 
\mathbf{A}_{m}^{\mathrm{H}} \mathbf{M} \mathbf{A}_{m}\right)-1=0,
\end{equation}

\begin{equation}
  \label{eq:29}
  \mathbf{R}_{2}=\sum_{m=1}^{H} m \mathrm{Im}
\left(\mathbf{A}_{m}\left(i_{norm}\right)\right)=0,
\end{equation}
where the $i_{norm}$ represents the DOF with phase constraint, to yield the extended residual equation
\begin{equation}
  \label{eq:30}
  \mathbf{R}(\hat{\mathbf{A}}, \omega_{nl}, \zeta_{nl}, k, a)=\left[\begin{array}{l}
    \mathbf{R}_{c} \\
    \mathbf{R}_{1} \\
    \mathbf{R}_{2}
    \end{array}\right]=\mathbf{0}.
\end{equation}
By exploiting a continuation technique, \emph{e.g.} arc-length method~\cite{Wang_ND,Wang_dNNM,Krack_book}, and fixing the wavenumber $k$ within the Irreducible Brillouin Zone, the nonlinear frequency, $\omega_{nl}$, and damping ratio, $\zeta_{nl}$, can be computed. Here, the code package {\it NLvib} 1.3~\cite{Krack_book} is used for the numerical continuation.

\section{Method of multiple scales}
\label{sec:A2}
The method of multiple scales (MMS)~\cite{RN30,RN31,RN32} requires the introduction of a book-keeping parameter, $\varepsilon$. Consequently, Eq.~\eqref{eq:1} is rewritten as:
\begin{equation}
    \label{eq:31}
    \mathbf{M}\ddot{\mathbf{x}}_n + \varepsilon^2\sum_{i=-1}^{1}{\mathbf{C}_i\dot{\mathbf{x}}_{n+i}}+ \sum_{i=-1}^{1}{\mathbf{K}_i\mathbf{x}_{n+i}} +\mathbf{F}_n\left(\mathbf{x}_{n}\right)+\mathbf{F}_{nb}\left(\mathbf{x}_{n}-\mathbf{x}_{n-1}\right) +\mathbf{F}_{nb}\left(\mathbf{x}_{n}-\mathbf{x}_{n+1}\right) = 0,
\end{equation}
where
\[
\begin{aligned}
\mathbf{M} = \begin{bmatrix}
m_h+m_a & m_a\\
m_a & m_a
\end{bmatrix},\mathbf{K}_{\pm1} = \begin{bmatrix}
    -\kappa_1 & 0\\
    0 & 0
\end{bmatrix},\mathbf{K}_0 = \begin{bmatrix}
    2\kappa_1 & 0\\
    0 && \kappa_2
\end{bmatrix}, 
\hat{\mathbf{C}}_{\pm1} = \begin{bmatrix}
    -\hat{c}_1 & 0\\
    0 & 0
\end{bmatrix},\\
\hat{\mathbf{C}}_0 = \begin{bmatrix}
    2\hat{c}_1 & 0\\
    0 & \hat{c}_2
\end{bmatrix}
\mathbf{x}_{n+i} = \begin{bmatrix}
    u_{n+i}\\
    v_{n+i}
\end{bmatrix}, 
\mathbf{F}_{nb} = \begin{bmatrix}
    \mp\varepsilon\hat{\beta}_1 (u_n-u_{n\pm1})^2 + \varepsilon^2\hat{\gamma}_1 (u_n-u_{n\pm1})^3\\
    0
\end{bmatrix}, \mathbf{F}_{n} = \begin{bmatrix}
    0\\
    \varepsilon\hat{\beta}_2 v_n^2 + \varepsilon^2\hat{\gamma}_2 v_n^3
\end{bmatrix},
\end{aligned}
\]
and the $\hat{c}_1 = c_1/\varepsilon^2, \hat{c}_2 = c_2/\varepsilon^2,\hat{\beta}_1=\beta_1/\varepsilon,\hat{\beta}_2=\beta_2/\varepsilon,\hat{\gamma}_1=\gamma_1/\varepsilon^2$, and $\hat{\gamma}_2 = \gamma_2/\varepsilon^2$. 

To solve Eq.~\eqref{eq:31}, the following assumptions are introduced:
\begin{equation}
\label{eq:32}
\begin{aligned}
\mathbf{x}_n &= \mathbf{x}_{n,0} + \varepsilon\mathbf{x}_{n,1} + \varepsilon^2\mathbf{x}_{n,2}+\cdots,\\
u_n &= u_{n,0}+\varepsilon u_{n,1} + \varepsilon^2 u_{n,2}+\cdots,\\
v_n &= v_{n,0}+\varepsilon v_{n,1} + \varepsilon^2 v_{n,2}+\cdots,\\
T_n &= \varepsilon^n t, d/dt = D_0 +\varepsilon D_1 +\varepsilon^2 D_2 + \cdots,\\
d^2/dt^2 &= D_0^2 + 2\varepsilon D_0 D_1 + \varepsilon^2(D_1^2 + 2D_0 D_2) + \cdots.
\end{aligned}
\end{equation}
Substituting these assumptions into Eq.~\eqref{eq:31} yields the following problems at different orders of $\varepsilon$:

\noindent\emph{1. Problem of order $\varepsilon^0$}
\begin{equation}
    \label{eq:33}
    \mathbf{M}D_0^2\mathbf{x}_{n,0} + \sum_{i=-1}^{1}{\mathbf{K}_i\mathbf{x}_{n+i,0}} = 0.
\end{equation}

\noindent\emph{2. Problem of order $\varepsilon$}
\begin{equation}
\label{eq:34}
    \mathbf{M}D_0^2\mathbf{x}_{n,1} + \sum_{i=-1}^{1}{\mathbf{K}_i\mathbf{x}_{n+i,1}}= - \begin{bmatrix}
        \hat{\beta_1}(u_{n,0}-u_{n-1,0})^2 - \hat{\beta_1}(u_{n,0}-u_{n+1,0})^2\\
        \hat{\beta_2} v_{n,0}^2
    \end{bmatrix} - 2\mathbf{M}D_0D_1\mathbf{x}_{n,0}.
\end{equation}

\noindent\emph{3. Problem of order $\varepsilon^2$}
\begin{equation}
\label{eq:35}
\begin{aligned}
    \mathbf{M}D_0^2\mathbf{x}_{n,2} + \sum_{i=-1}^{1}{\mathbf{K}_i\mathbf{x}_{n+i,2}} = - 2\begin{bmatrix}
        \hat{\beta_1}(u_{n,0}-u_{n-1,0})(u_{n,1}-u_{n-1,1}) - \hat{\beta_1}(u_{n,0}-u_{n+1,0})(u_{n,1}-u_{n+1,1})\\
        \hat{\beta_2} v_{n,0}v_{n,1}
    \end{bmatrix} \\
    - \begin{bmatrix}
         \hat{\gamma_1}(u_{n,0}-u_{n-1,0})^3 + \hat{\gamma_1}(u_{n,0}-u_{n+1,0})^3\\
        \hat{\gamma_2} v_{n,0}^3
    \end{bmatrix} 
    -\sum_{i=-1}^{1}{\mathbf{C}_iD_0\mathbf{x}}_{n+i,0}-\mathbf{M}\left(2D_0D_1\mathbf{x}_{n,1}+(D_1^2 + 2D_0D_2)\mathbf{x}_{n,0}\right).
\end{aligned}
\end{equation}

For the problem of order $\varepsilon^0$, the solution can be assumed as:
\begin{equation}
    \label{eq:36}
    \mathbf{x}_{n,0} = \frac{a}{2}\mathbf{\Phi}_{m,0}e^{j(\omega_{0} T_0-kn)} + c.c.,
\end{equation}
where $a$ is a complex variable satisfying $a = \rho_m e^{j\psi_m}$, with $\rho_m(T_1,T_2,\cdots)$ controlling the amplitude and $\psi_m(T_1,T_2,\cdots)$ describing the initial phase. The subscript $m$ indicates that these variables belong to MMS, distinguishing them from the results obtained using NF theory. Substituting Eq.~\eqref{eq:36} into Eq.~\eqref{eq:33} yields:
\begin{equation}
    \label{eq:37}
    \left(\mathbf{K}_0+2\cos{k}\mathbf{K}_{-1}-\omega_{0}^2\mathbf{M}\right)\mathbf{\Phi}_{m,0} = 0.
\end{equation}
Solving this eigenvalue problem provides the non-damping linear dispersion solutions, $\omega_{0}, \mathbf{\Phi}_{m,0}$, which can be used to denote the undamping acoustic or optic frequency and real wave shape. The eigenvectors $\mathbf{\Phi}_{m,0}$ satisfy the mass normalization, $\mathbf{\Phi}_{m,0}^{\mathrm{T}}\mathbf{M}\mathbf{\Phi}_{m,0} = 1$. Notably, when ignoring the damping in the metamaterial chain and focusing on the same branch, the relation $\mathbf{\Phi}_{1}=\frac{\mathbf{\Phi}_{m,0}}{\sqrt{2\lambda_{1}}}$ can be established by substituting it into the normalisation condition, $\mathbf{Y}_i^{\mathrm{T}}\mathbf{B}\mathbf{Y}_{s}=\delta_{is}$, and using $\mathbf{\Phi}_{m,0}^{\mathrm{T}}\mathbf{M}\mathbf{\Phi}_{m,0} = 1$. Consequently, the ratio of $\rho/\rho_m$ will follow $|\mathbf{\Phi}_{m,0}/\mathbf{\Phi}_{1}| = \sqrt{2\omega_{0}}$.

Substituting Eq.~\eqref{eq:36} into the problem of order $\varepsilon$ yields:
\begin{equation}
\label{eq:38}
\begin{aligned}
    \mathbf{M}D_0^2\mathbf{x}_{n,1} + \sum_{i=-1}^{1}{\mathbf{K}_i\mathbf{x}_{n+i,1}}&= - \begin{bmatrix}
        0\\
        \frac{\hat{\beta_2} a\bar{a}}{2}\mathbf{\Phi}_{m2,0}^2
    \end{bmatrix} -jD_1a \omega_{0} \mathbf{M} \mathbf{\Phi}_{m,0}e^{j(\omega_{0} T_0-kn)}\\
    &- \begin{bmatrix}
        \frac{\hat{\beta_1} a^2}{4}\mathbf{\Phi}_{m1,0}^2((1-e^{jk})^2-(1-e^{-jk})^2)\\
        \frac{\hat{\beta_2} a^2}{4}\mathbf{\Phi}_{m2,0}^2
        \end{bmatrix}e^{2j(\omega_{0} T_0-kn)} +c.c.
\end{aligned}
\end{equation}
where $\mathbf{\Phi}_{mi,0}$ represents the $i$-th element in $\mathbf{\Phi}_{m,0}$. The terms involving $e^{j(\omega_{0} T_0-kn)}$ are secular terms. To ensure solvability, these terms must be set to zero:
\begin{equation}
    \label{eq:39}
    D_1a = 0,
\end{equation}
After this operation, Eq.~\eqref{eq:38} becomes:
\begin{equation}
\label{eq:40}
\begin{aligned}
    \mathbf{M}D_0^2\mathbf{x}_{n,1} + \sum_{i=-1}^{1}{\mathbf{K}_i\mathbf{x}_{n+i,1}}&= - \begin{bmatrix}
        0\\
        \frac{\hat{\beta_2} a\bar{a}}{2}\mathbf{\Phi}_{m2,0}^2
    \end{bmatrix} \\
    &- \begin{bmatrix}
        \frac{\hat{\beta_1} a^2}{4}\mathbf{\Phi}_{m1,0}^2((1-e^{jk})^2-(1-e^{-jk})^2)\\
        \frac{\hat{\beta_2} a^2}{4}\mathbf{\Phi}_{m2,0}^2
        \end{bmatrix}e^{2j(\omega_{0} T_0-kn)} +c.c.
        \end{aligned}
\end{equation}
and with the solution of
\begin{equation}
    \label{eq:41}
    \mathbf{x}_{n,1} = a\bar{a}\mathbf{C}_{0,1} + a^2\mathbf{C}_{2,1}e^{2j(\omega_{0} T_0-kn)} + c.c.
\end{equation}
where
\[
\begin{aligned}
    \mathbf{C}_{0,1} &= -\begin{bmatrix}
    0\\
    \frac{\hat{\beta_2}}{2\kappa_2}\mathbf{\Phi}_{m2,0}^2
    \end{bmatrix},\\
    \mathbf{C}_{2,1} &= -(\mathbf{K}_0 + 2\cos{2k}\mathbf{K}_{-1}-(2\omega_{0})^2\mathbf{M})^{-1}\begin{bmatrix}
        \frac{\hat{\beta_1}}{4}\mathbf{\Phi}_{m1,0}^2((1-e^{jk})^2-(1-e^{-jk})^2)\\
        \frac{\hat{\beta_2}}{4}\mathbf{\Phi}_{m2,0}^2
    \end{bmatrix}.
\end{aligned}
\]

For the problem of order $\varepsilon^2$, substituting Eqs.~\eqref{eq:36} and~\eqref{eq:41} into Eq.~\eqref{eq:35} yields:
\begin{equation}
    \label{eq:42}
    \mathbf{M}D_0^2\mathbf{x}_{n,2} + \sum_{i=-1}^{1}{\mathbf{K}_i\mathbf{x}_{n+i,2}} = -N.S.T.e^{3j(\omega_{0} T_0-kn)} - S.T.e^{j(\omega_{0} T_0-kn)} + c.c.
\end{equation}
where $S.T.$ and $N.S.T.$ represent the secular and non-secular terms. And the $S.T.$ follows
\[
\begin{aligned}
S.T. &= \begin{bmatrix}
    a^2\bar{a}\hat{\beta_1}\mathbf{\Phi}_{m1,0}\mathbf{C}_{21,1}((1-e^{2jk})(1-e^{-jk})-(1-e^{-2jk})(1-e^{jk}))\\
    a^2\bar{a}\hat{\beta_2}\mathbf{\Phi}_{m2,0}\left(\mathbf{C}_{02,1} + \mathbf{C}_{22,1}\right)
\end{bmatrix}\\ 
&+\begin{bmatrix}
    \frac{3\hat{\gamma_1} a^2\bar{a}}{2}\mathbf{\Phi}_{m1,0}^3(1-\cos{k})^2\\
    \frac{3\hat{\gamma_2} a^2\bar{a}}{8}\mathbf{\Phi}_{m2,0}^3
\end{bmatrix}+j\omega_{0}(\mathbf{C}_0 + 2\cos{k}\mathbf{C}_1)\frac{a}{2}\mathbf{\Phi}_{m,0}+j\omega_{0}D_2a \mathbf{M}\mathbf{\Phi}_{m,0},
\end{aligned}
\]
where $\mathbf{C}_{0i,1}$ and $\mathbf{C}_{2i,1}$ implies the $i$-th elements in $\mathbf{C}_{0,1}$ and $\mathbf{C}_{2,1}$. Eliminating the secular term and substituting $a = \rho_m e^{j\psi_m}$ yields
\begin{equation}
    \label{eq:43}
    \begin{aligned}
    D_2\psi_m &= \frac{\rho_m^2\hat{\beta_1}\mathbf{\Phi}_{m1,0}\Re{(\mathbf{C}_{21,1}((1-e^{2jk})(1-e^{-jk})-(1-e^{-2jk})(1-e^{jk})))}}{\omega_{0}} \\
    &+ \frac{3\rho_m^2\hat{\gamma_1}\mathbf{\Phi}_{m1,0}^4(1-\cos{k})^2}{2\omega_{0}}+\frac{\rho_m^2\hat{\beta_2}\mathbf{\Phi}_{m2,0}^2(\mathbf{C}_{02,1}+\Re(\mathbf{C}_{22,1}))}{\omega_{0}} + \frac{3\rho_m^2\hat{\gamma_2}\mathbf{\Phi}_{m2,0}^4}{8\omega_{0}},\\
    D_2\rho_m &= -\frac{\mathbf{\Phi}_{m,0}^\mathrm{T}(\hat{\mathbf{C}}_0+2\cos{k}\hat{\mathbf{C}}_{-1})\mathbf{\Phi}_{m,0}\rho_m}{2}\\
    &-\frac{\rho_m^3\hat{\beta_1}\mathbf{\Phi}_{m1,0}^2\Im{(\mathbf{C}_{21,1}((1-e^{2jk})(1-e^{-jk})-(1-e^{-2jk})(1-e^{jk})))}}{\omega_{0}}\\
    &-\frac{\rho_m^3\hat{\beta_2}\mathbf{\Phi}_{m2,0}^2\Im(\mathbf{C}_{22,1})}{\omega_{0}}.
    \end{aligned}
\end{equation}
From these equations, the nonlinear frequency can be expressed as $\omega_{nl} = \omega_{0}+\varepsilon^2D_2\psi_m$, and the damping ratio is given by $\zeta_{nl} = - \varepsilon^2\frac{D_2\rho_m}{\omega_{nl}}$. After eliminating the secular terms, the $\mathbf{x}_{n,2}$ can be solved via the Eq.~\eqref{eq:42}
\begin{equation}
    \label{eq:44}
    \mathbf{x}_{n,2} = a^3\mathbf{C}_{3,2}e^{3j(\omega_{0} T_0-kn)}+c.c.
\end{equation}
where
\[
\begin{aligned}
\mathbf{C}_{3,2} &= -(\mathbf{K}_0+2\cos{3k}\mathbf{K}_{-1}-(3\omega_{0})^2\mathbf{M})^{-1}\left(\begin{bmatrix}
    \frac{\hat{\gamma_1}}{8}\mathbf{\Phi}_{m1,0}^3((1-e^{jk})^3+(1-e^{-jk})^3)\\
    \frac{\hat{\gamma_2}}{8}\mathbf{\Phi}_{m2,0}^3
\end{bmatrix}\right. \\
&\left.+\begin{bmatrix}
    \hat{\beta_1}\mathbf{\Phi}_{m1,0}\mathbf{C}_{21,1}((1-e^{2jk})(1-e^{jk})-(1-e^{-2jk})(1-e^{-jk}))\\
    \hat{\beta_2}\mathbf{\Phi}_{m2,0}\mathbf{C}_{22,1}
\end{bmatrix}\right).
\end{aligned}
\]
The invariant manifold can be expressed as $\mathbf{x}_n = \mathbf{x}_{n,0} + \varepsilon \mathbf{x}_{n,1} + \varepsilon^2 \mathbf{x}_{n,2}$.

\bibliographystyle{unsrt}

\bibliography{ref.bib} 
\end{document}